\documentclass[aip,cha,reprint]{revtex4-1}

\usepackage{amsmath}
\usepackage{amssymb}  % for fonts like \mathbb
\usepackage{bm}       % bloodbath symbols
\usepackage{booktabs} % defines top, middle and bottom rule in tables
\usepackage{graphicx} % for graphics
\usepackage{harpoon}  % \overrightharp to designate the transition rates
\usepackage{hyperref} % hyperlinks in the document
\usepackage{mathtools} % over bracket
\usepackage[usenames,dvipsnames]{xcolor} % used in hyperlinks 

\allowdisplaybreaks[1]

\newcommand{\apriori}{\textit{a priori}}
\newcommand{\ie}{\textit{i.e.}}    % id est
\newcommand{\eg}{\textit{e.g.}}    % exempli gratia
\newcommand{\etc}{etc}             % et cetera

\newcommand{\refcite}[1]{Ref.~[\onlinecite{#1}]}
\newcommand{\refcites}[1]{Refs.~\onlinecite{#1}}

\newcommand{\eq}[1]{(\ref{eq:#1})}
\newcommand{\eqlabel}[1]{\label{eq:#1}}
\def\eqreftwo(#1,#2){(\ref{eq:#1},\ref{eq:#2})}
\newcommand{\eqtwo}[1]{\eqreftwo(#1)}
\def\eqrefthree(#1,#2,#3){(\ref{eq:#1},\ref{eq:#2},\ref{eq:#3})}

\def\eqreffour(#1,#2,#3,#4){(\ref{eq:#1},\ref{eq:#2},\ref{eq:#3},\ref{eq:#4})}

\newcommand{\Fig}[1]{Fig.~\ref{fig:#1}}
\newcommand{\fig}[1]{fig.~\ref{fig:#1}}

\newcommand{\seclabel}[1]{\label{sec:#1}}
\newcommand{\secref}[1]{\ref{sec:#1}}

\newcommand{\Secn}[1]{Section~\secref{#1}}

\newcommand{\secn}[1]{Section~\secref{#1}}

\newcommand{\appn}[1]{Appendix~\secref{#1}}

\newcommand{\tablabel}[1]{\label{tab:#1}}
\newcommand{\tabref}[1]{\ref{tab:#1}}
\newcommand{\Tab}[1]{Table~\tabref{#1}}
\newcommand{\tab}[1]{Table~\tabref{#1}}

\newtheorem{defn}{Definition}

\newcommand{\second}{\mathrm{s}}
\newcommand{\mus}{\mu\second}
\newcommand{\ms}{\mathrm{ms}}

\ifx01
\ifx\notcolor\undefined\newcommand{\notcolor}{blue}\else\fi
\else
\ifx\notcolor\undefined\newcommand{\notcolor}{black}\else\fi
\fi
\newcommand{\+}[2]{\def#1{{#2}}}
\newcommand{\1}[2]{\def#1##1{{#2}}}
\newcommand{\2}[2]{\newcommand{#1}[2]{{#2}}}
\newcommand{\3}[2]{\newcommand{#1}[3]{{#2}}}

\+{\@}{\partial}                        % partial differential
\+{\bydef}{\;\triangleq\;}              % is by definition 
\+{\const}{\mathrm{const}}              % constant
\+{\dd}{\mathrm{d}}                     % ordinary differential
\2{\df}{\frac{\dd #1}{\dd #2}}          % ordinary derivative
\2{\ddff}{\frac{\dd^2 #1}{\dd #2^2}}    % second ordinary derivative
\1{\dom}{\mathop{\mathrm{dom}}\left(#1\right)} % domain of a function
\+{\i}{i}                               % \sqrt{-1}
\+{\krDelta}{\delta}                    % Kronecker delta
\1{\Mx}{\begin{bmatrix}#1\end{bmatrix}} % matrix defined by components
\+{\mnull}{\Matrix{0}}                  % null matrix
\1{\norm}{\left\lVert#1\right\rVert}    % norm VNB
\2{\pf}{\frac{\@{#1}}{\@{#2}}}          % partial derivative 
\2{\ppff}{\frac{\@^2{#1}}{\@{#2}^2}}    % second partial derivative
\3{\ppf}{\frac{\@^2 #1}{\@ #2 \@ #3}}   % mixed second partial derivative
\1{\Re}{\mathop{\mathrm{Re}}\left(#1\right)}      % real part of
\+{\R}{\mathbb{R}}                      % real set (TS)
\+{\Real}{\mathbb{R}}                   % real set (VNB)
\+{\T}{^T}                              % transposed
\2{\takenat}{\left.#1\rule[-2ex]{0pt}{4ex}\right|_{#2}}
\1{\ve}{\Vector{e}_{#1}}                % unit vector
\+{\vi}{\Vector{1}}                     % vector of unities
\+{\vnull}{\Vector{0}}                  % null vector 
\+{\vnullT}{\vnull\T\ }                 % null vector -- vertical

\newcommand{\Matrix}[1]{\hat #1}                 % matrix
\newcommand{\Vector}[1]{\mathbf{#1}}              % column-vector
\newcommand{\gVector}[1]{\bm{#1}}                 % column-vector, Greek

\+{\st}{t}                    % standard time
\+{\stmax}{\st_{\max}}        % simulation duration
\+{\ft}{\tau}                 % fast time t = eps*tau
\+{\delt}{\Delta\st}          % time step TS
\+{\dt}{\Delta_t}             % time step VNB
\1{\ddt}{\df{#1}{\st}}        % time derivative
\1{\ddtt}{\ddff{#1}{\st}}     % second time derivative
\1{\ddft}{\df{#1}{\ft}}       % fast-time derivative
\1{\dfat}{\df{#1}{\at}}       % autonomous time derivative, ordinary
\1{\pfat}{\pf{#1}{\at}}       % autonomous time derivative, partial
\+{\uu}{u}                    % a dynamical variable
\+{\vu}{\Vector{\uu}}         % vector of dynamical variables
\+{\vf}{\Vector{f}}           % unperturbed vector field
\+{\vh}{\Vector{h}}           % vector field perturbation
\+{\eps}{\varepsilon}         % small parameter epsilon
\+{\epsi}{\frac{1}{\eps}}     % inverse of a small parameter epsilon
\1{\cO}{\mathcal{O}(\eps^{#1})}

\+{\N}{n}                     % dimensionality of the system/maximal component index (TS)
\+{\m}{m}                     % multiplicity of zero eigenvalue -- dimensionality of the manifold
\+{\n}{n}                     % dimensionality of the system/maximal component index (VNB)

\+{\Ei}{\mathrm{i}}           % Markov state enumerator i
\+{\Ej}{\mathrm{j}}           % Markov state enumerator j
\+{\iin}{i}                   % index from 1..n
\+{\iinj}{j}                  % index from 1..n
\+{\iini}{j_1}                % index from 1..n
\+{\iinii}{j_2}               % index from 1..n
\+{\iim}{k}                   % index from 1..m
\+{\iimi}{p}                  % index from 1..m
\+{\imn}{\ell}                % index from m+1..n
\+{\itn}{q}                   % index from 2..n
\+{\itni}{q'}                 % index from 2..n
\+{\itm}{r}                   % index from 2..m
\+{\itmi}{r'}                 % index from 2..m
\+{\iat}{0}                   % index of eigenvalue of autonomous time
\+{\imc}{1}                   % starting index of Markov chain
\+{\imin}{\iota}              % 0/1, starting value of the eval enumerator

\+{\U}{U}                     % the equilibrium manifold components
\+{\vU}{\Vector{\U}}          % .., vector
\+{\V}{v}                     % solution perturbation component
\+{\vv}{\Vector{\V}}          % .., vector
\+{\A}{a}                     % coordinates on the manifold component
\+{\va}{\Vector{\A}}          % .., vector

\+{\B}{b}                     % manifold perturbation component
\+{\vb}{\Vector{\B}}          % .., vector

\+{\at}{\sigma}               % autonomous time
\+{\mF}{\Matrix{F}}           % Jacobian of the vector field
\1{\eval}{\Lambda_{#1}}        % Jacobian eigenvalue
\+{\evec}{V}                  % Jacobian right eigenvector component
\1{\vevec}{\Vector{\evec}_{#1}} % .., vector
\1{\vadvec}{\Vector{W}_{#1}}  % Jacobian left eigenvector
\1{\vadvecT}{\Vector{W}_{#1}^T} % .., transposed (row-vector) 
\+{\x}{x}                     % state probability
\+{\vx}{\Vector{\x}}          % .., vector
\+{\mA}{\Matrix{A}}           % transition matrix
\1{\Ai}{A_{#1}}               % .., components
\+{\mAf}{\Matrix{A}_f}        % .., fast part
\+{\mAfT}{\Matrix{A}_f^T}     % .., .., transposed
\+{\mAs}{\Matrix{A}_s}        % .., slow part of transition matrix
\+{\tmA}{\tilde{\Matrix{A}}}  % .., rescaled
\1{\tAi}{\tilde{A}_{#1}}      % .., .., components
\1{\vevecA}{\gVector{\kappa}_{#1}} % right evector of the fast matrix
\1{\vevecAT}{\vevecA{#1}^T}  % .., transposed
\+{\mevecA}{\Matrix{K}}       % .., matrix of
\2{\evecA}{\kappa^{#1}_{#2}}  % .., component of
\1{\vadvecA}{\gVector{\rho}_{#1}} % left evector of the fast matrix
\1{\vadvecAT}{\vadvecA{#1}^T} % .., transposed 
\+{\madvecA}{\Matrix{P}}      % .., matrix of
\2{\advecA}{\rho^{#1}_{#2}}   % .., component of
\1{\evalA}{\lambda_{#1}}      % eigenvalue of Af
\+{\evalAcc}{\lambda^*}       % .., its complex conjugate
\+{\Pseudoinverse}{\Matrix{L}}% matrix of the inversion in the stable subspace

\+{\vUx}{\vU^\x}              % equilibrium manifold, Markov part
\+{\vvx}{\vv^\x}              % .., correction to
\1{\vevecx}{\Vector{\evec}^{\x}_{#1}} % Jacobian right eigenvector, Markov part
\+{\vevecT}{\mu}              % Jacobian right eigenvector, t-part

\+{\mM}{\Matrix{M}}           % transition matrix of the new system
\+{\mMl}{\Matrix{M}_0}        % .., leading order
\+{\mMf}{\Matrix{M}_1}        % .., first order correction

\+{\vm}{V_m}                  % transmembrane voltage
\+{\Naion}{\mathrm{Na}}       % sodium 
\+{\ina}{I_\Naion}            % sodium current
\+{\gna}{g_\Naion}            % ..., max conductivity
\+{\ena}{E_\Naion}            % ..., reversal potential
\+{\I}{I}                     % other currents
\+{\l}{\ell}                  % ..., counting index.
\+{\vX}{\Vector{X}}           % other dynamic variables

\+{\ifB}{\cal{I}}             % integrating factor
\+{\iifB}{\cal{J}}            % integrating factor exponential
\+{\iv}{\xi}                  % integrating variable
\+{\ivi}{s}                   % integrating variable

\+{\lint}{k}                  % intermediate integrating limit
\+{\lit}{\st_n}               % upper integrating limit
\+{\lil}{s_\imn}              % lower time integrating limit

\+{\iC}{C_\imn}               % arbitrary constant
\+{\iD}{D_\imn}               % C*I

\+{\fO}{\mathcal{F}}       % first-order term
\+{\vfO}{\mathcal{\Vector{\fO}}} % vector first-order term
\+{\Gdyn}{\mathcal{G}}     % a term with derivative inside first-order
\+{\lO}{\mathcal{L}}       % leading-order term

\1{\HH}{\mathcal{H}_\imn(#1)} % free term in the b-eqn
\+{\BBl}{B_0}                 % main term in the solution of the b-eqn
\+{\BBf}{B_1}                 % perturbation term in the solution of the b-eqn
\+{\tBBl}{\tilde{B}_0}        % something when solving b-eqn
\+{\tBBf}{\tilde{B}_1}        % something when solving b-eqn

\+{\mH}{\Matrix{H}}               % Jacobian matrix of (d h/d u)

\1{\TR}{\trate{#1}}             %transition rates, shorter version
\2{\TT}{#1\cdot#2}              % reciprocal transition rates
\1{\trate}{\overset{\rightharpoonup}{#1}} %transition rates
\1{\ftrate}{\beta_{#1}}         % rate fractions in OP reduction
\1{\STUtrate}{\gamma_{#1}}         % rate fractions in STU reduction
\1{\QRtrate}{\delta_{#1}}         % transition rates in the first order correction

\+{\xO}{O}
\+{\xP}{P}
\+{\xQ}{Q}
\+{\xR}{R}
\+{\xS}{S}
\+{\xT}{T}
\+{\xU}{U}
\+{\xV}{V}
\+{\xW}{W}

\+{\aN}{{\tilde{N}}}
\+{\aQ}{{\tilde{Q}}}
\+{\aU}{\tilde{U}}
\+{\aR}{R}
\+{\aS}{S}
\+{\aT}{T}
\+{\aV}{V}
\+{\aW}{W}

\+{\aM}{\tilde{M}}
\+{\aL}{\tilde{L}}

\+{\aNref}{\tilde{N}_{\mathrm{ref}}}

\+{\diagO}{D_{\xO}}               % {-(\trate{OP}+\trate{OU})}
\+{\diagP}{D_{\xP}}               % {-(\trate{PQ}+\trate{PU}+\trate{PO})}
\+{\diagQ}{D_{\xQ}}               % {-(\trate{QR}+\trate{QT}+\trate{QP})}
\+{\diagR}{D_{\xR}}               % {-(\trate{RS}+\trate{RQ})}
\+{\diagS}{D_{\xS}}               % {-(\trate{ST}+\trate{SR})}
\+{\diagT}{D_{\xT}}               % {-(\trate{TQ}+\trate{TS}+\trate{TU})}
\+{\diagU}{D_{\xU}}               % {-(\trate{UT}+\trate{UP}+\trate{UO}+\trate{UV})}
\+{\diagV}{D_{\xV}}               % {-(\trate{VU}+\trate{VW})}
\+{\diagW}{D_{\xW}}               % {-\trate{WV}}

\+{\fON}{\fOtrate{\aN}}      % first-order term

\+{\mMs}{\Matrix{M}_s}                % transition matrix of the new system -- slow transitions
\+{\mMfa}{\Matrix{M}_f}                % transition matrix of the new system -- fast transitions

\+{\alRQ}{\alpha_{RQ}}
\+{\alQP}{\alpha_{QP}}
\+{\alPO}{\alpha_{PO}}
\+{\alST}{\alpha_{ST}}
\+{\alTU}{\alpha_{TU}}
\+{\alQR}{\alpha_{QR}}
\+{\alPQ}{\alpha_{PQ}}
\+{\alOP}{\alpha_{OP}}
\+{\alTS}{\alpha_{TS}}
\+{\alUT}{\alpha_{UT}}
\+{\alUP}{\alpha_{UP}}
\+{\alTQ}{\alpha_{TQ}}
\+{\alSR}{\alpha_{SR}}
\+{\alPU}{\alpha_{PU}}
\+{\alQT}{\alpha_{QT}}
\+{\alRS}{\alpha_{RS}}
\+{\alOU}{\alpha_{OU}}
\+{\alUO}{\alpha_{UO}}
\+{\alUV}{\alpha_{UV}}
\+{\alVU}{\alpha_{VU}}
\+{\alVW}{\alpha_{VW}}
\+{\alWV}{\alpha_{WV}}
\+{\alUN}{\alpha_{UN}}
\+{\alQN}{\alpha_{QN}}
\+{\alNU}{\alpha_{NU}}
\+{\alNQ}{\alpha_{NQ}}
\+{\alNO}{\alpha_{NO}}
\+{\alNP}{\alpha_{NP}}
\+{\alMS}{\alpha_{MS}}
\+{\alMT}{\alpha_{MT}}
\+{\alMU}{\alpha_{MU}}
\+{\alRM}{\alpha_{RM}}
\+{\alQM}{\alpha_{QM}}
\+{\alPM}{\alpha_{PM}}
\+{\alVM}{\alpha_{VM}}
\+{\alOM}{\alpha_{OM}}
\+{\alMO}{\alpha_{MO}}
\+{\alMP}{\alpha_{MP}}
\+{\alMQ}{\alpha_{MQ}}
\+{\alMR}{\alpha_{MR}}
\+{\alMV}{\alpha_{MV}}
\+{\alLR}{\alpha_{LR}}
\+{\alLQ}{\alpha_{LQ}}
\+{\alPL}{\alpha_{PL}}
\+{\alLP}{\alpha_{LP}}
\+{\alML}{\alpha_{ML}}
\+{\alLM}{\alpha_{LM}}

\+{\DF}{*}
\begin{document}
\title{Fast-slow asymptotics for a Markov chain model of fast sodium current} %Title of paper
\author{Tom\'a\v{s} Star\'y, Vadim N. Biktashev}
\affiliation{%
  College of Engineering, Mathematics and Physical Sciences,
  University of Exeter, 
  Harrison Building, 
  North Park Road, 
  Exeter EX4 4QF, 
  United Kingdom
}
\date{\today}

\begin{abstract}
  We explore the feasibility of using fast-slow asymptotic to
  eliminate the computational stiffness of the discrete-state,
  continuous-time deterministic Markov chain models of ionic channels
  underlying cardiac excitability. We focus on a Markov chain model of
  the fast sodium current, and investigate its asymptotic behaviour
  with respect to small parameters identified in different ways.
\end{abstract}

% \tableofcontents


\maketitle 

\begin{quotation}
  Modern models of excitability of cardiac cells include description of
  ionic channels in terms of deterministic Markov chains. The
  transition rates in these chains often vary by several orders of
  magnitude, which makes numerical simulations more difficult and
  necessitates development of specialized numerical approaches. We
  follow the usual wisdom that the small parameters in a mathematical
  model can be turned from an impediment into an advantage by
  development of asymptotic description exploiting those small
  parameters. The usual problem with experiment-derived (rather than
  postulated) models is that small parameters in them are not defined
  a priori but need to be identified, and sometimes this can be done
  equally plausibly in more than one way. In this paper, we show how the
  standard fast-slow asymptotic theory can be applied to the Markov chain
  models of ionic channels using one important model of this class as an
  example. We explore three selected ways of identifying small parameters in
  this model, and investigate the factors on which the utility of resulting
  asymptotics depends. 
\end{quotation}

\section{Introduction}
\seclabel{intro}

The bioelectricity is one of the driving forces of our life. Our
mind and body is a manifestation of complex dynamics of electric
impulses that carry information and trigger reactions in different
organs of our body. The pulses of electrical excitation in heart are responsible
for starting a chain of reactions, resulting in contraction of the
cardiac muscle, causing the blood circulation. Understanding the
detailed mechanisms of formation and propagation of electrical
excitation can help in treatment and prevention of cardiac diseases. 

The direct experimentation with living systems is difficult and
rises many ethical issues, hence a mathematical description provides
a valuable tool to gain insight and understanding of the internal
working of the heart. 

The elementary part of a cardiac excitation model are models of the
ion-specific channel in the membrane, that close or open in response to the
change in the transmembrane voltage. On the molecular scale, functioning of a
single channel is an inherently stochastic process, which is adequately
described as continuous time Markov chain (or Markov processes, as they are
sometimes called). For most applications, however, the stochastic component is
not essential, and it is sufficient to describe the behaviour of the channel
in terms of the deterministic ``master equation'' for the probabilities of the
channel to be in certain states, as functions of time. Simulation of resulting
excitation models for single cells does not create problems; but when scaled
to the tissue or whole-organ level, this becomes computationally expensive
\cite{Plank-etal-2008}.  The computer
  technology is constantly improving, and recent publications describe results
  that would be unthinkable a few years before: \eg\
  \textcite{Richards-etal-2013} simulated excitation in whole human heart with
  spatial resolution of 0.13\,mm in nearly real time.  However, that was done
  on a system with 1.6 million CPUs and a peak speed of 20 petaflops. For the
  moment, such systems are far from ubiquitous. It is therefore important
to try and improve the computational methods for simulation of cardiac
excitation models.

One significant factor of computational complexity is that the Markov chain
models of ionic channels often involve processes on time scales differing by
several orders of magnitude, i.e. are stiff. So a direct approach using
explicit time steppers requires very small time steps, hence high
computational demands. For instance,
  \textcite{Bondarenko-2014}, for a model involving a number of Markov chain 
  channels, used time steps as short as 2 picoseconds---this is to be
  compared to the duration of the onset of an action potential of the order of
  1 millisecond, and duration of the cardiac pulse of the order of 1 second.
The natural alternative is generic implicit time steppers. However, approaches
based on exploiting specific properties of cardiac excitation model present an
attractive third possibility. In this paper, we explore one possible way to
exploit specific properties of the Markov chain models of ionic channels. This
is based on the traditional idea that a small parameter in the model can be
turned from a hindrance into an advantage, by finding asymptotics in this
parameter. For small parameters responsible for numerical stiffness, the
adequate approach is singular perturbation theory, of fast-slow asymptotics.

The problem of stiffness of the description of the ion
  gates' dynamics is not new for the Markov chain models, and was there
  already for the Hodkin-Huxley (HH) type ``gate'' models, starting from the
  \textcite{Hodgkin-Huxley-1952}. Out of many approaches to overcome stiffness
  in these models, arguably the most popular one follows the work by
  \textcite{Rush-Larsen-1978}. In some respects, it offers and ``ideal''
  solution for the HH-type models, combining accuracy and stability.  This
  approach exploits the fact that the equation controlling an HH-type gate is
  quasi-linear, in the sense that is is linear with respect to the gate
  variable, although it depends on other, ``control'', variables in a
  non-linear way. Typically, the control variable is the transmembrane
  voltage, but in some channels it is concentration of Ca ions in addition or
  instead of the transmembrane voltage. If the control variables change
  negligibly during one time step, then ``freezing'' these variables allows one
  to write an ``exact'' solution.  In the simplest formulation, the resulting
  computational scheme is first-order accurate, but the coefficient in the
  leading order term depends on the rate of the voltage (or whatever the
  control variable is) and does not depend on the stiffness of the equation
  with respect to the gate variable. This approach may be be formalized as
  fast-slow approach, where the control variable is slow and gate variable
  is fast. In that case, the leading-order solution for the gate variable,
  the ``instant equilibrium'', corresponds to the limit when the time step is
  much longer than the characteristic time constant defined by the current
  values of the transition rate. However, the Rush-Larsen scheme in fact
  retains its accuracy when the time step is much less than, or is comparable
  to this characteristic time.

The Rush-Larsen approach has been so popular, it
  inevitably spawned a number of attempts to improve and/or extend
  it. For instance, \textcite{Perego-Veneziani-2009} proposed how to
  increase its accuracy from first to second order, and
  \textcite{Marsh-etal-2012} suggested how to apply it to equations
  which are not quasi-linear, by linearising them for the duration of
  the time step. The Markov chain description poses a different sort
  of challenge: the equations are quasi-linear with respect to the
  Markov state occupancies so there is no need in linearization, but
  instead of one equation for a gate variable, there is a system of
  simultaneous linear equations. The ``straightforward''
  Rush-Larsen-type approach, when each of the equations of the system
  is considered in turn, by freezing all other variables together with
  the control variables, has been proven effective for a number of
  examples (see \refcite{Plank-etal-2008} and references therein).  A
  more radical approach for Markov chain generalization of the
  Rush-Larsen scheme, which utilized the exact solution for the whole
  linear system, thus avoiding extra errors caused by freezing some
  Markov states while updating others, was described
  in~\refcites{Stary2015,Stary-Biktashev-2015-CinC}.  

  The methods discussed above offer efficient numerical schemes, but do not
  exploit the fact that the different processes within the same
  Markov chain may, and often do, have vastly different speeds. Taking these
  into account can, at least theoretically, offer further advantages.  There
  have been a number of inspiring examples of this kind in literature.  For
  instance, \textcite{Hinch-etal-2004} and \textcite{Plank-etal-2008} used the
  concept of ``rapid equilibrium'', exploiting rapid transition rates between
  some of the states of a Markov chain, to effectively ``merge'' the closely
  connected states into one ``combined'' state, for a number of Markov chain
  models. This approach can be formalized as a leading-order asymptotic in the
  classical fast-slow perturbation theory descending from the works by
  \textcite{Tikhonov-1952} and \textcite{Fenichel-1979}, for a particular form
  in which a small parameter $\eps$ appears in the equations: as a factor
  $1/\eps$ in front of the transition rates between two selected Markov
  states.

        In the present paper, we seek to analyse this sort of
        asymptotics in more detail, following the full formalism,
        rather than immediately getting to the answer by following the
        rather obvious, but still only intuitive ``rapid equilibrium''
        argument. The motivation for a more detailed analysis includes
        possibilities of generalization of the asymptotic approach to
        the fast/slow separation cases other than pairs of fast
        reciprocal transition rates, and getting higher-order terms in
        asymptotics. Moreover, an important theoretical question is
        whether stiff Markov chain formulations of ionic channels can
        always be well described by the standard singular perturbation
        theory. The intrigue here comes from the fact that the
        Hodgkin-Huxley description of some cardiac channels,
        specifically the fast sodium current channel, may be only
        partly described by the Tikhonov asymptotics, in regards of
        the the activation, `$m$'-gate; whereas the attempts to treat
        the inactivation, `$h$'-gate as fast or slow compared to the
        transmembrane voltage are ineffective, and at the very least
        fail to describe some essential qualitative features, such as
        the maximum of the action potential which is lower than the
        reversal potential for the sodium ions and which is different
        in a single cell in a propagating wave in tissue, to name the
        simplest
        example~\cite{%
          Biktashev-2002-PRL,%
          Biktashev-Suckley-2004-PRL,%
          Biktasheva-etal-2006-PTRSA,%
          Simitev-Biktashev-2011}.
        Hence the answer for the Markov chain model of the same
        channel is far from obvious \apriori\ and requires
        investigation.   

          To address the theoretical question posed above, we have
          chosen the Markov chain model of the fast sodium current
          developed by \textcite{Clancy2002}. This is not the stiffest
          model of existing models of the kind, but it is one of the
          most popular ones. As our study was methodological rather
          than practical, it was important for our choice that the
          transition rates, and related characteristic times, in this
          model are varied in wide ranges which makes identification
          of small parameters a nontrivial issue.  In other words, our
          aim was not to identify examples when the standard
          asymptotics can successfully treat the problem of stiffness
          (such examples do exist, see above), but rather analyse
          cases when the standard approach fails, as a necessary step
          towards developing more adequate, non-standard approaches.
          An extra motive for the choice of the fast Na current in
          this context was that the Hogdkin-Huxley description of this
          current is known to require non-Tikhonov asymptotics, as
          discussed above.

The structure of the paper is as follows. \Secn{dim-red} introduces
the notation and main principles of the singular perturbation approach
we are using. \Secn{mc-red} discusses amendments required of this
approach with account of the specifics of Markov chain models.
\Secn{embedding} presents a formalization of the process of
identification of small parameters in experiment-based models, which
we call parametric embedding. \Secn{app-ina} introduces the Markov
chain model of the fast sodium current which we use to apply the
singular perturbations.  \Secn{emb-ina} presents the main results,
coming out of a few different parametric embeddings of this
model. This is concluded by discussion in~\secn{discussion}. We also
present an Appendix containing technical material which is required
for reproducing the main results but not for their understanding.

\section{General Theory for Dimensionality Reduction}
\seclabel{dim-red}

The singular perturbation theory is well known in a variety of
different formulations. We mostly follow the terminology and notation used
e.g. in~\refcites{Biktashev2003,Suckley2003,Biktashev2004,Biktasheva2006,Biktashev2008},
adjusting where necessary for our present purposes. 

We consider an autonomous system of ordinary differential equations
\begin{align} \eqlabel{ode-def}
  \ddt{\vu} = \vf(\vu) + \eps\vh(\vu)
\end{align}
where $\vu,\vf,\vh \in \R^\n$, and $\eps$ is a small positive
parameter.
We assume existence of 
a stable $\m$-dimensional manifold $\{\vU\}$ of equilibria of the unperturbed
system, $\eps=0$, 
i.e. $\vf(\vU)=\vnull$, where $\vnull$ stands for the null vector, 
with coordinates $\va\in\R^\m$, $1\le\m<\n$, and 
looking for solutions of the perturbed system, $\eps>0$, in the form
\begin{align} \eqlabel{ansatz-solution}
  \vu = \vU(\va) + \eps \vv(\st)
\end{align}
where the perturbation of the solution $\vb \in \R^\n$ is orthogonal
to the manifold, in the sense that 
\begin{align} \eqlabel{solution-perturbation}
  \vv(\st) = \sum_\imn \B_\imn \vevec\imn(\va) ,
\end{align}
where the vectors $\vevec\iin(\va)$ are right eigenvectors of a Jacobian
matrix $\mF(\vU) = \@\vf/\@\vu|_{\vu = \vU}$, 
\begin{align} \eqlabel{F-eproblem}
  \mF \vevec\iin = \eval\iin \vevec\iin,
\end{align}
and the summation index $\imn$ runs through the stable eigenvalues,
$\Re{\eval\imn}<0$, $\imn=\m+1,\dots,\n$, skipping the zero eigenvalues,
$\eval\iim$, $\iim=1\,\dots,\m$, corresponding to the directions tangent to
the manifold.  \Tab{eig-indices} summarises the meaning of these and other
index conventions as used throughout the text, subject to a small amendment in
the next section.

\begin{table}[b]\tablabel{eig-indices}
\caption{%
  Ranges of indices used in the text, unless explicitly stated otherwise. %
}
\begin{tabular}{lll}
  \toprule
  \textbf{index}            & \textbf{values}    & \textbf{corresponds to} \\
  \midrule
  $\iin,\iinj,\iini,\iinii$ & $\imin,\hdots,\N$ \footnotemark[1] & all eigenvalues of Jacobian (J.) \\
  $\iim,\iimi$              & $\imin,\hdots,\m$ \footnotemark[1] & zero eigenvalues of J. \\
                            & $\iat$            \footnotemark[2] & autonomous time direction \\
  $\itn,\itni$              & $\imc,\hdots,\N$  \footnotemark[2] & all eigenvalues of Markov chain (M.C.) \\    
  $\itm,\itmi$              & $\imc,\hdots, \m$ \footnotemark[2] & zero eigenvalues of M.C. $\evalA\itm = 0$ \\    
  $\imn$                    & $\m+1,\hdots,\N$  & non-zero eigenvalues of J. or M.C. \\
  \bottomrule
\end{tabular} 
\footnotetext[1]{
  Here $\imin=1$ for~\secn{dim-red} and $\imin=0$ from the next
  section on.} 
\footnotetext[2]{
  This is used starting from \secn{mc-red}.} 
\end{table}

The right eigenvectors corresponding to zero eigenvalues
$\eval\iim = 0$ are tangent to the invariant manifold and can be
found as
\begin{align}
  \label{eq:evec-tang}
  \vevec\iim = \pf{\vU}{\A_\iim}.
\end{align}

We substitute \eq{ansatz-solution} and \eq{solution-perturbation} into
\eq{ode-def}, expand the nonlinear functions into their Taylor series and
separate the components using left eigenvectors $\vadvecT{\iin}$ as
projectors. The detailed derivation is presented in the \appn{reduct-comp}.
The final result reads as the following system of ODEs
\begin{subequations} \eqlabel{split-ode}
    \begin{align}
      \label{eq:split-ode-a}
    \epsi\ddt{\A_\iim} 
    =&
       \vadvecT{\iim}\vh(\vU) +
       \eps \fO_\iim(\va,\vb)+
       \cO{2}, \\
      \label{eq:split-ode-b}
    \ddt{\B_\imn}
    =&
       \eval\imn\B_\imn  +
       \vadvecT{\imn}\vh(\vU) +
       \cO{},
  \end{align}
\end{subequations}
where in the right-hand side of the first equation we have kept the
leading order term, $\vadvecT{\iim}\vh(\vU)$, and the first-order
correction $\fO_\iin$, which works out as
\begin{align} \eqlabel{first-order}
  \fO_\iin(\va,\vb) = 
  & \vadvecT{\iin} \mH(\vU) \vv
    +
    \vadvecT{\iin}\sum_{\iini,\iinii}\ppf{\vf}{\uu_\iini}{\uu_\iinii} \V_\iini\V_\iinii 
  \nonumber \\ 
  & \mbox{} + \epsi\sum_{\iim}     \pf{\vadvecT{\iin}}{\A_\iim }\ddt{\A_\iim }\vv ,
\end{align}
where
\begin{align}
  \label{eq:jacob-H}
      \mH(\vU) = \takenat{\pf{\vh}{\vu}}{\vu = \vU} .
\end{align}

Equations~\eq{split-ode-a} and \eq{split-ode-b}  are
coupled through higher-order terms, and to complete the reduction, we
need to eliminate $\vb$. 
For the solution of the manifold
coordinates $\va$ up to $\cO{N}$ it is sufficient to find the
correction term $\vb$ up to $\cO{N-1}$. 
The leading order term for the correction $\vb$ in terms of $\va$ can
be found by solving \eqref{eq:split-ode-b} using the integrating
factor method.  The solution also requires Taylor expansion of the
integrating factor and of the non-homogeneous term. This leads to
\begin{align} \eqlabel{corr-solution-gen}
  \B_\imn = -\frac{\vadvecT{\imn}\vh(\vU)}{\eval\imn} + \cO{},
\end{align}
which is to be substituted into the first-order term into
\eqref{eq:split-ode-a}, which then becomes a closed equation for
$\va$.

\section{Dimensionality Reduction for Time-Inhomogeneous Markov Chains}
\seclabel{mc-red}

The master equation for Markov chain models of ionic channels can be
written in the form
\begin{align}
  \label{eq:mc-ode}
  \ddt{\vx} = \mA(\st) \vx.
\end{align}
Entries in the vector of dynamical variables $\vx\in\Real^\n$ represent the
probabilities, that an ion channel resides in a particular
state. Entries of the transition matrix $\mA\in\Real^{\n\times\n}$ describe the conditional
probabilities of a channel in one given state to transit to another
state per unit of time, i.e. transition rates. In reality, the matrix
$\mA$ depends on other dynamic variables of the model, e.g. the
transmembrane voltage, which in turn are affected by the dynamics of
the Markov chain; however this is not essential for the formalism we
describe here and we assume that $\mA$ is an explicit function of
time, just for simplicity of notation. 
The sum of the entries in the vector of dynamical variables is equal
to $1$, \ie\ it is a \emph{stochastic vector}. 
This implies that the sum of the entries in each column of
the transition matrix $\mA$ has to be equal to $0$. This is achieved
as the entries on the diagonal of the transition matrix are a sum of
the entries out of the diagonal for each column of the matrix. This
property together with the fact that the non-diagonal elements are
non-negative constitutes the definition of $\mA$ as a \emph{left-stochastic
matrix}.

To use the theory described in \secn{dim-red} we have to take into
account one simplifying fact and two complications.  The simplifying
fact is that the system \eq{mc-ode} is linear.  The complications are,
firstly, that the theory described in the previous section applies to
an autonomous system, but the Markov chain in \eqref{eq:mc-ode} has an
explicit time dependence of the transition matrix
$\mA(\st)$. Secondly, the theory requires a small parameter, however
the Markov chain models contain transition rates determined
experimentally, and identifiction of any small parameters in such a
case is a separate task, sometimes nontrivial.

The first complication is dealt with using autonomisation, which means
that we introduce an additional dynamical variable $\at$ to represent time
(henceforth referred to as \emph{``autonomous time''}). 
Then the vector of dynamical variables is
\[
  \vu=\Mx{\at \\ \vx}
\]
and the dynamic equation is
\begin{align}
  \label{eq:autonomisation}
  \ddt{} \Mx{ \at \\ \vx } = \Mx{ 1 \\ \mA(\at) \vx} .
\end{align}
Note that system \eq{autonomisation} is no longer
  linear unless the function $\mA(\at)$ is a constant.

To address the second complication, we introduce the small parameters
artificially in an empirical procedure we call parametric embedding,
which is discussed in detail in the next section. For now it is
important that as a result, we can split the transition rates matrix
$\mA$ into a fast part $\mAf$ and a slow part $\mAs$, and the
difference between them is identified by the small parameter $\eps$
appearing as
\begin{align}
  \label{eq:embedding}
  \mA = \epsi\mAf + \mAs.
\end{align}
We restrict consideration to the embeddings in which $\mAf$ and $\mAs$
are left-stochastic matrices. 
We assume that the fast matrix $\mAf(\at)$ is diagonalizable, and 
introduce the eigenvalues $\evalA\itn(\at)$ and the
right eigenvectors $\vevecA\itn(\at)$:
\begin{align}
  \label{eq:eval-mA}
  \mAf(\at)\vevecA\itn(\at) =& \evalA\itn(\at) \vevecA\itn(\at) 
\end{align}
(and drop from now on the dependence on $\at$, for brevity).
We assume that for all $\at$, matrix $\mMf$ has a full set of eigenvectors, 
the first $\m\ge1$ of the eigenvalues are
zero, and the remaining are all real (and of course
negative)\footnote{%
  Note that diagonalizability and reality of the eigenvalues of the
  full transition rate matrix $\mA$ can be guaranteed under the
  assumption of detailed balance~\cite{detailed-balance},
  and $\mAf=\lim_{\eps\to0}\left(\eps\mA\right)$. %
}. Correspondingly, we introduce also the left eigenvectors
$\vadvecA\itn$,
\begin{align*}
  \mAfT\vadvecA\itn = \evalA\itn \vadvecA\itn ,
  \qquad
  \vadvecAT\itn\vevecA\itni = \krDelta_{\itn,\itni}. 
\end{align*}
Differentiation of the last identity with respect to $\at$ yields a
relationship that will be useful: 
\begin{align}
  \label{eq:dynamic-kron-mc}
  \dfat{\vadvecAT{\iin}}\vevecA\iinj =   -\vadvecAT{\iin}\dfat{\vevecA\iinj}.
\end{align}

We transform the system \eqref{eq:autonomisation} into fast time
$\ft = \st/\eps$ to get a system
\begin{align}
  \label{eq:aut-emb-mc}
  \ddft{\at} =& \eps
               , \\
  \nonumber
  \ddft{\vx} =& \mAf(\at)\vx + \eps\mAs(\at)\vx
               ,
\end{align}
This can be considered in the format of \eq{ode-def} with
\begin{align}
  \label{eq:mc-into-generic}
  \vu =
  \begin{bmatrix}
    \at  \\ \vx
  \end{bmatrix},
  \qquad
  \vf =
  \begin{bmatrix}
    0
    \\
    \mAf(\at)\vx
  \end{bmatrix} ,
  \qquad 
    \vh =
  \begin{bmatrix}
    1
    \\
    \mAs(\at)\vx
  \end{bmatrix}.
\end{align}
The dimensionality of the autonomized system \eq{aut-emb-mc} is
$\n+1$; we keep the upper value of the corresponding indices as $\n$
but reserve the value $\iat$ for the time variable $\at$; this is
where parameter $\imin$, designating the minimal value of the
  eigenvalues' indices in ~\tab{eig-indices}, becomes $0$.

The manifold of equilibria in this case is in fact a linear subspace
of $\Real^{\n+1}$ which is the hull of the one-dimensional subspace
corresponding to the time coordinate $\at$ and the kernel of the fast
matrix $\mAf$:
\begin{align} \eqlabel{lin-mfd}
  \vU(\at,\va) = \Mx{ \A_\iat \\  \vUx(\va) }
\end{align}
where
$\A_\iat=\at$, $\va=\Mx{\A_1,\dots\A_\m}\T$ and
\begin{align} \eqlabel{ansatz-solution-x}
  \vUx(\va) =   
  \sum_\itm \A_\itm\vevecA\itm.
\end{align}

To construct the reduced system, we need to find the Jacobian of function
$\vf$ and solve the eigenvalue problem. The Jacobian is easily found as
\begin{align}
  \label{eq:eval-jacob}
    \mF(\vU) =
  \begin{bmatrix}
    0 & \vnullT
    \\
    \dfat{\mAf}\vUx & \mAf 
  \end{bmatrix}.
\end{align}
Let us denote the components
of the eigenvectors as
\begin{align}
  \label{eq:eval-evecs}
  \vevec\iin =
  \begin{bmatrix}
    \vevecT_\iin
    \\
    \vevecx\iin
  \end{bmatrix} .
\end{align}
Substituting \eq{eval-jacob} and \eq{eval-evecs} into
$\eval\iin\vevec\iin=\mF\vevec\iin$, we get 
\begin{subequations}
  \begin{align}
    \label{eq:eval-jacob-at}
    \eval\iin \vevecT_\iin &=0,
    \\
    \label{eq:eval-jacob-Af}
    \eval\iin \vevecx\iin&=\dfat{\mAf}\vUx \vevecT_\iin + \mAf\vevecx\iin.
  \end{align}
\end{subequations}
Let us consider separately the cases $\vevecT_\iin=0$ and $\vevecT_\iin\ne0$.
For $\vevecT_\iin=0$, equation \eqref{eq:eval-jacob-Af} becomes
\begin{align}
  \label{eq:eval-mA-1}
  \mAf\vevecx\itn =& \eval\itn \vevecx\itn ,
\end{align}
which is the definition of an eigenvalue problem for matrix $\mAf$,
so we can take $\vevecx\itn=\vevecA\itn$ and $\evalA\itn =
\eval\itn$, for $\itn=\imc,\dots,\N$, out of
which the first $\m$ are zero eigenvalues.
For $\itm = \imc,\hdots,\m$, we have $\eval\itm=0$, and
differentiation of \eqref{eq:eval-mA} with respect to $\at$ gives
\begin{align}
  \label{eq:zero-eval-derive}
   \dfat{\mAf}\vevecA\itm = - \mAf\dfat{\vevecA\itm}.
\end{align}

We find one more eigenpair for the case $\vevecT_\iat \neq 0$.  Let us
normalise the corresponding eigenvector so that $\vevecT_\iat=1$.
Then to satisfy equation \eqref{eq:eval-jacob-at} we must have
$\eval\iat=0$, and \eqref{eq:eval-jacob-Af} becomes
\begin{align}
  \label{eq:eval-jacob-Af-at}
  \mAf\vevecx\iat= -  \dfat{\mAf}\vUx .
\end{align}
If we substitute \eqref{eq:ansatz-solution-x} into
\eqref{eq:eval-jacob-Af-at} and use \eqref{eq:zero-eval-derive}, we
get
\begin{align}
  \mAf \vevecx\iat =  \mAf\sum_\itm \A_\itm\dfat{\vevecA\itm} 
\end{align}
so we can choose
\begin{align}
  \label{eq:vevecA-iat}
  \vevecA\iat =   \sum_\itm \A_\itm\dfat{\vevecA\itm}
\end{align}
(this is choice is of course non-unique because the zero
eigenvalue has multiplicity $\m+1$). 

The left eigenvectors are treated similarly. To summarise the results,
the eigenvalues $\eval\iin$ and eigenvectors $\vevec\iin$, $\vadvec\iin$ of the Jacobian in the time-extended
system are related to those $\evalA\itn$, $\vevecA\itn$, $\vadvecA\itn$ of the transition rate matrix via the
following relationships: 
\begin{align}
  \label{eq:evec-jacob-mA}
  &\vevec\iat =
  \begin{bmatrix}
    1
    \\
    \sum_\itm \A_\itm\dfat{\vevecA\itm}
  \end{bmatrix},&&
  \vevec\itn =
  \begin{bmatrix}
    0
    \\
    \vevecA\itn
  \end{bmatrix},
  \\ \nonumber
  &\vadvec\iat =
  \begin{bmatrix}
    1
    \\
    \vnull
  \end{bmatrix},&&
  \vadvec\itn =
  \begin{bmatrix}
    -\vadvecAT{\itn}     \sum_\itm \A_\itm\dfat{\vevecA\itm}
    \\
    \vadvecA\itn
  \end{bmatrix}%  =
  , \\
  &\eval0 = 0, && \eval\itn = \evalA\itn . \nonumber
\end{align}
With these the time-component of the ``leading-order term'' works out
as $\vadvecT{\iat }\vh(\vU) =1$, as should be expected.
For the Markov chain subspace, we use
\eq{dynamic-kron-mc}, \eq{mc-into-generic}, \eq{ansatz-solution-x} and \eq{evec-jacob-mA} to get
\begin{align*}
  &
  \vadvecT{\itn}\vh(\vU) = 
  \Mx{ \dfat{\vadvecAT{\itn}} \vUx, & \vadvecAT\itn}
  \Mx{ 1 \\ \mAs(\at)\vx }
  \\ &
  =  -\vadvecAT{\itn}  \sum_\itm  \A_\itm \left( \dfat{\vevecA\itm}
  - \mAs(\at) \vevecA\itm \right) ,
\end{align*}
and then 
\eq{corr-solution-gen} gives
the formula for the components of the transversal correction,
\begin{align}
  \label{eq:sol-across}
  \B_\imn =
  \frac{\vadvecAT{\imn}  }{\evalA\imn}    \sum_\itm \A_\itm \left( \dfat{\vevecA\itm}
  -
  \mAs(\at)\vevecA\itm \right)+
       \cO{} ,
\end{align}
and the transversal correction itself as
\begin{align}
  \label{eq:pert-vvx}
  \vvx = \sum_\imn \B_\imn     \vevecA\imn =
  \sum_{\imn,\itm}   \left[ \frac{      \A_\itm  }{\evalA\imn}\vadvecAT{\imn}\left(    \dfat{\vevecA\itm}
    -
    \mAs(\at)\vevecA\itm \right)\right]
    \vevecA\imn.
\end{align}
Finally, the first-order accurate reduced system of ODEs is given by
\eqref{eq:split-ode-a}, leading to
\begin{subequations}
\begin{align}
  \label{eq:split-ode-a-expand-1}
  &\epsi\ddt{\A_\iat} 
    =
    1        + \cO{2} ,\\
  \label{eq:a-itm-diff}
  &\epsi\ddt{\A_\itm} 
    =
    \left(\dfat{\vadvecAT{\itm}}
    +
    \vadvecAT{\itm}
    \mAs(\at)\right) \sum_\itmi \A_\itmi\vevecA\itmi 
  \\ \nonumber 
  &+ \eps   \left(\vadvecAT{\itm} \mAs(\at) +
   \dfat{\vadvecAT{\itm}}
  \right)  \\ \nonumber
  &\times \sum_{\imn,\itmi}  \left[  \frac{ \A_\itmi }{\evalA\imn} \vadvecAT{\imn}\left(    \dfat{\vevecA\itmi}
    -
    \mAs(\at)\vevecA\itmi \right)\right]
    \vevecA\imn
       + \cO{2}.
\end{align}
\end{subequations}
This result can be written in the matrix form as
\begin{align} \eqlabel{red-M}
  \epsi\ddt\va % = \mM\va 
  = \left( \mMl(\at) + \eps \mMf(\at) \right) \va + \cO{2},
\end{align}
where
\begin{align} \eqlabel{red-Ml}
  \mMl = \left[ \madvecA'+\madvecA\mAs \right]\,\mevecA,
\end{align}
\begin{align}  \eqlabel{red-Mf}
  \mMf & = \left[ \madvecA'+\madvecA\mAs \right]
  \Pseudoinverse
  \left[ \mevecA' - \mAs\mevecA \right],
\end{align}
\begin{align} \eqlabel{red-KPL}
  \mevecA(\at) & = \Mx{\vevecA1 | \dots | \vevecA\m} = \Mx{
    \evecA11 & \dots  & \evecA1\m \\
    \vdots    & \ddots & \vdots \\
    \evecA\n1 & \dots & \evecA\m\n
  }
  \in \Real^{\n\times\m},
  \nonumber \\
  \madvecA(\at) & = \Mx{ \vadvecAT1 \\ \vdots \\ \vadvecAT\m }=\Mx{
    \advecA11 & \dots  & \advecA\n1 \\
    \vdots    & \ddots & \vdots \\
    \advecA1\m & \dots & \advecA\n\m
  }
  \in\Real^{\m\times\n}, 
  \nonumber \\ 
  \Pseudoinverse(\at) & =
  \sum_{\imn} \vevecA\imn \left(\evalA\imn\right)^{-1} \vadvecAT\imn 
  \in\Real^{\n\times\n}. 
\end{align}
and dash $'$ stands for differentiation with respect to $\at$.

\section{Parametric embedding}
\seclabel{embedding}

To address the second complication, we introduce the small parameters
artificially in a procedure known as parametric embedding,
previously introduced
in~\refcites{Suckley2003,Biktasheva2006,West-etal-2015-JMB}. 
This procedure is a formalization of the
replacement of a small constant with a small parameter. 

\begin{defn}
  We will call a system %
  \[
  \dot{u} = F(u;\eps), \qquad u\in\Real^{d},
  \]
  depending on parameter $\eps$, 
  a \emph{one-parametric embedding} of a system
  \[
  \dot{u} = f(u), \qquad u\in\Real^{d},
  \]
  if $f(u)\equiv F(u,1)$ for all $u\in\dom{f}$. 
  If the limit $\eps\to0$ is concerned then we call it an
  \emph{asymptotic embedding}. 
\end{defn}

The typical use of this procedure has the form of a replacement of a
small constant with a small parameter. If a system contains a
dimensionless constant $a$ which is ``much smaller than 1'', then
replacement of $a$ with $\eps a$ constitutes a 1-parametric
embedding; and then the limit $\eps\rightarrow0$ can be
considered. In practice, constant $a$ would more often be replaced
with parameter $\eps$ rather than multiplied by it, but
mathematically speaking, in the context of $\eps\rightarrow0$ and
$a=\const\ne0$, these two ways are formally equivalent. \emph{This explains
the paradoxical use of a zero limit for a parameter whose true value
is one.}

In some applications, the ``small parameters'' appear naturally and
are readily identified. However, this is not always the case, and in
complex systems identification of adequate small parameters may be a
task in itself, which is where the formalization of this procedure can
be helpful. In the context of the definition above, it is important to
understand that there are infinitely many ways a given system can be
parametrically embedded, as there are infinitely many ways to draw a
curve $F(u;\eps)$ in the functional space given the only
constraint that it passes through a given point, $F(u;1)=f(u)$.  In
terms of asymptotics, which of the embeddings is ``better'' depends on
the qualitative features of the original systems that need to be
represented, or classes of solutions that need to be
approximated. 

If a numerical solution of the system can be found easily, then there
is a simple practical recipe: to look at the solutions of the
embedding at different, progressively decreasing values of the
artificial small parameter $\eps$, and see when the features of
interest will start to converge. If the convergent behaviour is
satisfactorily similar to the original system with $\eps=1$, the
embedding is adequate for these features.  

To summarize, we claim that identification of small parameters in a given
mathematical model with experimentally measured functions and constants will,
from the formal mathematical viewpoint, always be arbitrary, even though in
the simplest cases the choice may be so natural that that this ambiguity is
not even realized by the modeller, and that ``validity'' of such
identification can be defined only empirically: if the asymptotics describe
the required class of solutions sufficiently well. The exceptions may be when
the asymptotic series are in fact convergent, the approximation errors can be
estimated \apriori, but this is rare.

In the subsequent text, slightly abusing the above definition for the
sake of brevity, we refer as ``embedding'' to particular instances of
one-parametric embedding of a given system for a selected value of the
parameter $\eps$. The overall structure of the embeddings is always
like in~\eq{embedding}, and the difference is in the choice of the matrices
$\mAf$ and $\mAs$. 

\section{Definition of the Markov Chain model of the fast sodium
  current (INa)}
\seclabel{app-ina}

We apply the asymptotic theory described above to a Markov chain model of the
fast sodium current developed by \citet{Clancy2002} (we consider the wild-type
version). The relevant part of the model has the form
\begin{subequations} \eqlabel{fullmodel}
\begin{align}
    \ddt{\vm} & = - \gna \left[ \vm - \ena(\vX) \right] \xO - \sum_\l \I_\l(\vX),   \\
    \ddt{\vx} & = \mA(\vm) \vx, \eqlabel{submodel} \\
    \ddt{\vX}  & = \dots
  \end{align}
\end{subequations}
where $\vm$ is the transmembrane voltage, $\gna$ is the maximal conductance
of the fast Na current, $\ena$ is the reversal potential of the Na ions due
to the transmembrane difference in the concentration of these ions, $\xO$ is
one of the components of the vector $\vx$ representing the fraction of open
channels, corresponding to the open state of the fast Na current channels,
$\I_\l$ represent all other transmembrane currents, and the vector $\vX$
comprises all other dynamic variables of the model, such as other ionic
channels, concentrations etc.

\begin{figure}
  \centering
  \includegraphics{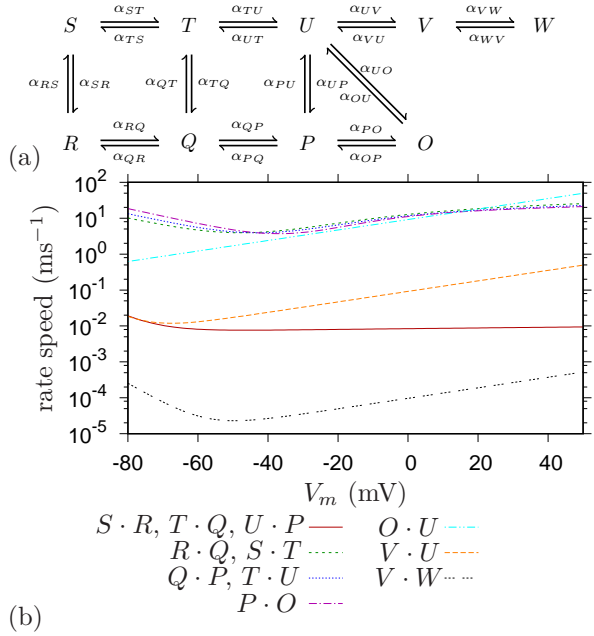}
  \caption{(a) Diagram of $\ina$ channel and (b) speed of transition
    rates in the range of physiological cell membrane voltages.
    Here $\TT SR=\TR{SR}+\TR{RS}$ etc are
      reciprocal transition rates, defining the speed with which the
      two given states tend to equilibrate with each other. 
  }
  \label{fig:diag-tr}
\end{figure}

\Fig{diag-tr}(a) shows the diagram
of the Markov chain. We find it more convenient to rename the dynamic
variables, \ie\ the names of the states of the Markov chain, as
reported in \refcite{Stary2015}: these are single-letter names, as
opposed to the original names in~\refcite{Clancy2002} which use up to
three symbols.  The only state in the model that corresponds to the
the channel being open is $\xO$, and this name coincides with the
nomenclature used by~\citeauthor{Clancy2002}.  So, for this Markov
chain we have $\N=9$ and the state vector
\[
  \vx = [\xO, \xP, \xQ, \xR, \xS, \xT, \xU, \xV, \xW]\T .
\]
According to the diagram of~\fig{diag-tr}(a), the transition rate
matrix has the structure
\begin{align}
  \label{eq:mA-inax}
  \mA =& 
  \begin{bmatrix}
\DF        &\trate{PO} &0          &0          &0          &0          &\trate{UO} &0          &0\\
\trate{OP} &\DF        &\trate{QP} &0          &0          &0          &\trate{UP} &0          &0\\
0          &\trate{PQ} &\DF        &\trate{RQ} &0          &\trate{TQ} &0          &0          &0\\
0          &0          &\trate{QR} &\DF        &\trate{SR} &0          &0          &0          &0\\
0          &0          &0          &\trate{RS} &\DF        &\trate{TS} &0          &0          &0\\
0          &0          &\trate{QT} &0          &\trate{ST} &\DF        &\trate{UT} &0          &0\\
\trate{OU} &\trate{PU} &0          &0          &0          &\trate{TU} &\DF        &\trate{VU} &0\\
0          &0          &0          &0          &0          &0          &\trate{UV} &\DF        &\trate{WV}\\
0          &0          &0          &0          &0          &0          &0          &\trate{VW} &\DF
  \end{bmatrix}.
\end{align}
Here and elsewhere in transition rates matrices, in the interests of saving space, we
do not show diagonal elements and replace them with $\DF$: they are
uniquely defined by the condition that the sum of elements in each row should
vanish. So \eg\ the top left diagonal element in the above matrix is
$-\trate{PO}-\trate{UO}$, and the bottom right element is $-\trate{VW}$.

All the transition rates in $\mA$ are functions of the transmembrane voltage
$\vm$. Their exact definitions can be found in the original
publication~\cite{Clancy2002} (see also
  \refcites{Stary2015,Stary-2016}) and we do not present them here;
however~\fig{diag-tr}(b) gives a graphical illustration of the magnitudes of
these rates in the physiological range of $\vm$. In that figure, we use the
sum of the transition rates between two states as the measure of the speed of
their connection, \ie\
$\TT\Ei\Ej\bydef\trate{\Ei\Ej}+\trate{\Ej\Ei}$. Indeed, it
is this quantity that determines the speed with which the dynamic
equilibrium between the two states is reached if occupancies of all other
states are fixed.

\section{Embeddings of the INa Model}
\seclabel{emb-ina}

\Fig{diag-tr}(b) allows one to see what transition rates may be
considered ``fast'' and thus included into $\mAf$. For
  instance, connections $PO$, $QP$, $TU$ are relatively fast in the
  whole range of voltages, connection $VW$ is always slow, whereas
  connection $OU$ is fast at high $\vm$ but not so high at low $\vm$,
  and connection $VU$ is somewhat intermediate between the group of
  clearly fast connection and the group of clearly slow
  connections. In accordance with the above discussed formal
  definition and informal semantics of the concept of embedding, we
  intend to treat the question of which connections can or should be
  considered fast as strictly empirical, so \Fig{diag-tr}(b) does
  not provide the ultimate answer to this question, but merely the
  possible directions of search.  For simplicity, we always construct
$\mAf$ by including into it reciprocal pairs of transition rates, and
completing the diagonal elements to ensure $\mAf$ is
left-stochastic. As a result, any nonzero non-diagonal element of
$\mA$ is always included into $\mAf$ and/or into $\mAs$, which
guarantees that $\mAf$ and $\mAs$ are left-stochastic.
We assess the quality of an embedding by how well it
  approximates the transients of the Markov states in a typical
  solution, a standard action potential; and of course of all the
  states the most important is the open state $\xO$. 

\begin{figure*}
  \centering
  \includegraphics{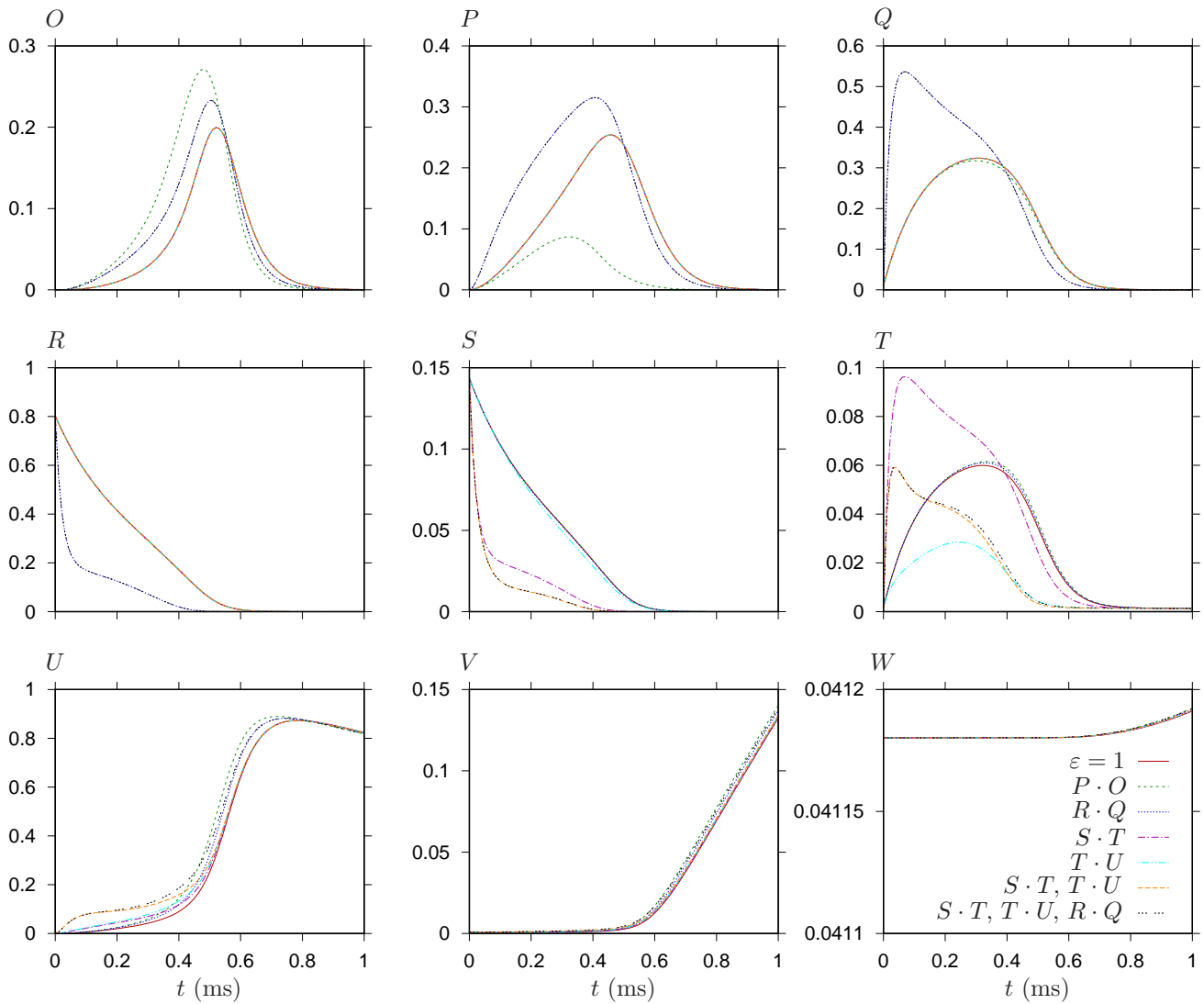}
  \caption[]{%
    Time evolution of state occupancy of $\ina$ with transition
    rates embeddings. The vertical axis shows the occupancy of the
    states from $O$--$W$ in alphabetical order as specified in top
    left corner. The red lines show the original model ($\eps = 1$),
    other lines show the embeddings as specified in the legend
      of panel $W$, \eg\      
    the green lines show the embeddings of transition rates between
    states $OP$, \ie\ both
    $\TR{OP}$ and $\TR{PO}$
    for $\eps=0.1$, 
    the grey line shows the embedding of the reciprocal transition rates
    between $ST$, $TU$ and $RQ$, etc.
  }
  \label{fig:embeddings}
\end{figure*}

We have tried a number of different combinations of reciprocal
  transition rates for $\mAf$. Not all such combinations pass the
  embedding test, \ie\ give reasonable approximation of the original
  solution in the limit $\eps\to0$. In particular, the would-be
  ``straightforward'' solution to consider as ``fast'' all the
  transition rates that appear so in~\fig{diag-tr}(b), does not
  work\cite{Stary-2016}.  Figure~\ref{fig:embeddings} shows results of
  simulation of some of the more successful of those
combinations. The $\ina$ model was extracted from the authors code
\cite{Clancy2002}. The simulation of the model were driven by
recorded values of $\vm(\st)$ during a standard action
potential from a single-cell simulation. 
That means, we have performed a simulation of the
  original full model~\eq{fullmodel} once, and the resulting function
  $\vm(\st)$ was then used for computations of the embedded version of only
  the subsystem ~\eq{submodel}, 
  in which $\vm(\st)$ was considered given and fixed; in other words,
  performed ``virtual voltage clamp'' experiments.
The time step in the simulation of $\ina$ was $\dt=1\;\mus$. The
original model is shown with red lines, the transition rates
embeddings are shown for a value of $\eps = 0.1$.

As can be deduced from the figure, although the transition rates
included in the embeddings have roughly the same orders of magnitude,
their expected effect on the accuracy of approximation of the $\xO$
transient by asymptotic methods is rather different: the $\xO\xP$
embedding is relatively poor, the $\xR\xQ$ is somewhat better, whereas
embeddings involving transitions between $\xS$, $\xT$ and $\xU$, any
pair or all three, promises very good accuracy: the corresponding
graphs are indistinguishable in the plot
resolution. Note that this assessment heavily depends
  on the special role of the $\xO$ state, and would be completely
  different if we were more interested in another Markov state. For
  instance, for the $\xS(\st)$ transient, the OP embedding promises
  good accuracy, and STU embedding is very poor. Obviously, it matters
  how close are the embedded rates to the state in question.

\subsection{OP-embedding}

In this section we develop an example of a particular embedding of the
transition rates between the states $\xO$ and $\xP$, \ie\ rates $\trate{OP}$
and $\trate{PO}$. As seen from the above discussion, the empirical evidence
suggests that the asymptotics of this embedding is not likely to give a good
approximation, so the purpose of this exercise is mainly didactic, to
demonstrate in detail the application of the general theory, including the
first-order correction, on a simple example.

In this embedding, the transition matrix $\mA$ is split according to
\eqref{eq:embedding} into the matrix of the slow transition rates 
\begin{align}
  \label{eq:mAs-ina}
    \mAs =& 
  \begin{bmatrix}
\DF         &0                        &0          &0          &0          &0          &\trate{UO} &0          &0\\
0           &\DF                      &\trate{QP} &0          &0          &0          &\trate{UP} &0          &0\\
0           &\trate{PQ}               &\DF        &\trate{RQ} &0          &\trate{TQ} &0          &0          &0\\
0           &0                        &\trate{QR} &\DF        &\trate{SR} &0          &0          &0          &0\\
0           &0                        &0          &\trate{RS} &\DF        &\trate{TS} &0          &0          &0\\
0           &0                        &\trate{QT} &0          &\trate{ST} &\DF        &\trate{UT} &0          &0\\
\trate{OU}  &\trate{PU}               &0          &0          &0          &\trate{TU} &\DF        &\trate{VU} &0\\
0           &0                        &0          &0          &0          &0          &\trate{UV} &\DF        &\trate{WV}\\
0           &0                        &0          &0          &0          &0          &0          &\trate{VW} &\DF \\
  \end{bmatrix}
\end{align}
and the matrix of the fast transition rates
\begin{align}
  \mAf =& 
  \begin{bmatrix}
\DF        &\trate{PO} &0 &0 &0 &0 &0 &0 &0\\
\trate{OP} &\DF        &0 &0 &0 &0 &0 &0 &0\\
0          &0          &0 &0 &0 &0 &0 &0 &0\\
0          &0          &0 &0 &0 &0 &0 &0 &0\\
0          &0          &0 &0 &0 &0 &0 &0 &0\\
0          &0          &0 &0 &0 &0 &0 &0 &0\\
0          &0          &0 &0 &0 &0 &0 &0 &0\\
0          &0          &0 &0 &0 &0 &0 &0 &0\\
0          &0          &0 &0 &0 &0 &0 &0 &0\\
  \end{bmatrix}.
\end{align}

For the dimensionality reduction we need to calculate the eigenvalues
and eigenvectors of the fast matrix $\mAf$. This will result to a
number of zero eigenvalues corresponding to the zero part of the
matrix. There will be also at least one zero eigenvalue
$\evalA\imc = 0$ corresponding to the Markov chain since
\begin{align}
  \label{eq:zero-eval}
  \vi\T \mA = 0 .
\end{align}
In fact, we have $\evalA\itm=0$ for $\itm=1,\dots,\m$, where $\m=8$,
and just one non-zero eigenvalue $\evalA9=
-(\trate{PO}+\trate{OP})$. The corresponding right eigenvectors are 
\begin{align} \eqlabel{rev-OP}
  \vevecA{\imc} &= (\trate{PO} + \trate{OP})^{-1}\left(
                    \trate{PO}\,\ve1+\trate{OP}\,\ve2\right),
                   \nonumber \\
  \vevecA{\iin} &= \ve{\iin+1}, \qquad \iin=2,\dots,8, \\\nonumber
  \vevecA{9} &= -\ve1+\ve2, \nonumber 
\end{align}
where $\ve\iin$ is the standard notation for the column-vector which has $\iin$-th component equal
to one and all other components equal to zero, 
% \begin{align} \eqlabel{rev-OP}
%   % \input{evec-mAfx.tex},
%   \vevecA{\imc} =&(\trate{PO} + \trate{OP})^{-1}
%                     \Mx{\trate{PO},&\trate{OP},&0,&0,&0,&0,&0,&0,&0}\T, \nonumber \\
%   \vevecA{2} =&\Mx{0,&0,&1,&0,&0,&0,&0,&0,&0}\T , \\\nonumber
%   \vevecA{3} =&\Mx{0,&0,&0,&1,&0,&0,&0,&0,&0}\T , \\\nonumber
%   \vevecA{4} =&\Mx{0,&0,&0,&0,&1,&0,&0,&0,&0}\T , \\\nonumber
%   \vevecA{5} =&\Mx{0,&0,&0,&0,&0,&1,&0,&0,&0}\T , \\\nonumber
%   \vevecA{6} =&\Mx{0,&0,&0,&0,&0,&0,&1,&0,&0}\T , \\\nonumber
%   \vevecA{7} =&\Mx{0,&0,&0,&0,&0,&0,&0,&1,&0}\T , \\\nonumber
%   \vevecA{8} =&\Mx{0,&0,&0,&0,&0,&0,&0,&0,&1}\T , \\\nonumber
%   \vevecA{9} =&\Mx{-1 ,&1,&0,&0,&0,&0,&0,&0,&0}\T , \nonumber 
% \end{align}
%
and the left eigenvectors are
\begin{align}  \eqlabel{OPlev}
  \vadvecA1 &= \ve1+\ve2, \\\nonumber
  \vadvecA{\iin} &=  \ve{\iin+1}, \qquad \iin=2,\dots,8, \\\nonumber
  \vadvecA9 &= (\trate{PO}+\trate{OP})^{-1} % \\\nonumber & \times 
      \left(-\trate{OP}\,\ve1+\trate{PO}\,\ve2\right).
\end{align}
% \begin{align}  \eqlabel{OPlev}
%   \vadvecA1 =& \Mx{1,&1,&0,&0,&0,&0,&0,&0,&0}\T, \\\nonumber
%   \vadvecA{\iin} =& \vevecA{\iin}, \qquad \iin=2,\dots,8, \\\nonumber
%   \vadvecA9 =& (\trate{PO}+\trate{OP})^{-1} \\\nonumber
%               & \times \Mx{-\trate{OP},&\trate{PO},&0,&0,&0,&0,&0,&0,&0}\T.
% \end{align}
%
We note that the left eigenvector $\vi$ asserted by the
identity~\eq{zero-eval} is a linear combination of these, namely
$\vi=\sum_{\iin=1}^8\vadvecA\iin$. The choice of normalization for
$\vevecA1$ and $\vadvecA1$ is motivated by the ease of
interpretation of the slow variable $\A_1$, which will transpire
shortly below.

Now we are ready to substitute the specifics of the selected embedding
into the equation \eqref{eq:a-itm-diff} describing the reduced model.
The left eigenvectors are constant for all $\itm$, so their
derivatives are zero, and $\{\imn\} = \{9\}$.  Then upon substituting
\eq{sol-across} into \eq{a-itm-diff} we get
\begin{align}  \label{eq:ina-a-itm-diff}
  \epsi\ddt{\A_\itm} = &
    \vadvecAT{\itm} \mAs(\at) \sum_\itmi\A_\itmi\vevecA\itmi 
  \\ \nonumber &
    + \eps\left[\vadvecAT{\itm} \mAs(\at)\vevecA9\B_9\right]
    + \cO{2}.
\end{align}
The differential equation for $\A_\itm$ for $\itm = 3,4,5,7,8$ come
out identical to the equations for the states $\xR,\xS,\xT,\xV,\xW$
from \eqref{eq:mA-inax}. This is because the first-order term vanishes
as $\vadvecAT{\itm} \mAs(\at) \vevecA9 =0$ for these $\itm$. Hence we
retain the same names for the corresponding components of the reduced
model, as they had in the original model, and the vector of dynamic
variables in the reduced system has the form
\begin{align}
  \va = [\aN,\aQ,\aR,\aS,\aT,\aU,\aV,\aW]\T,
\end{align}
where $\aN\bydef\A_1$, $\aQ\bydef\A_2$ and $\aU\bydef\A_6$. 

The components $\itm =1,2,6$ in \eqref{eq:ina-a-itm-diff}, that is
differential equations for $\A_1 = \aN$, $\A_2 = \aQ$ and $\A_6 = \aU$, will
have nonzero first-order terms.  According to $\vadvecA1$ as given by
\eq{OPlev}, the new variable $\aN$ is just a sum of the old states occupancies
$\xO$ and $\xP$; this is where the chosen normalization for $\vadvecA1$ comes
helpful.  The names of the slow variables $\aQ$ and $\aU$ are motivated by the
fact that according to~\eq{OPlev} they map exactly to $\xQ$ and $\xU$
respectively, and the difference from the old variables is only in the
first-order corrections in the reduced differential equations they obey.

Equation
\eq{ansatz-solution-x} then
defines the relationship between the original and the reduced variables
in the leading order, which in our case is 
\begin{align} \eqlabel{projection-leading}
  \vUx 
  = \Mx{\ftrate{PO}\aN,&\ftrate{OP}\aN,&\aQ,&\aR,&\aS,&\aT,&\aU,&\aV,&\aW},
\end{align}
where we define the fractions of the transition rates as
\begin{align}
  \ftrate{\Ei\Ej} = \frac{\trate{\Ei\Ej}}{\trate{PO} + \trate{OP}} .
\end{align}
We have only one stable eigenvalue in the present case, so equation
\eq{solution-perturbation} reduces to
\begin{align} \eqlabel{vvx-OP}
  \vvx = \B_9 \vevecA{9} ,
\end{align}
and equation \eqref{eq:sol-across},  with account of $\vevecA\itmi'=0$,
$\itmi\neq1$, gives
\begin{align} \eqlabel{sol-across-ina}
  \B_9 = \frac{\vadvecAT{9}}{\evalA9}
  \left( \A_1\vevecA1' - \mAs(\at)\vUx \right).
\end{align}
Then the leading-order transition matrix, according to \eqtwo{red-Ml,red-KPL}
is
\begin{align}
  & \mMl = \\ &
  \begin{bmatrix}
\DF           & \trate{QP} & 0          & 0          & 0          & (\trate{UO}+\trate{UP}) & 0          & 0          \\
\trate{\aN Q} & \DF        & \trate{RQ} & 0          & \trate{TQ} & 0                       & 0          & 0          \\
0             & \trate{QR} & \DF        & \trate{SR} & 0          & 0                       & 0          & 0          \\
0             & 0          & \trate{RS} & \DF        & \trate{TS} & 0                       & 0          & 0          \\
0             & \trate{QT} & 0          & \trate{ST} & \DF        & \trate{UT}              & 0          & 0          \\
\trate{\aN U} & 0          & 0          & 0          & \trate{TU} & \DF                     & \trate{VU} & 0          \\
0             & 0          & 0          & 0          & 0          & \trate{UV}              & \DF        & \trate{WV} \\
0             & 0          & 0          & 0          & 0          & 0                       & \trate{VW} & \DF     
  \end{bmatrix}, \nonumber
\end{align}
where the new transition rates are defined as
\begin{align}
  \trate{\aN Q} =& \ftrate{OP}\trate{PQ},\\ \nonumber
  \trate{\aN U} =& \ftrate{PO} \trate{OU}+\ftrate{OP}\trate{PU},
\end{align}
and the first-order correction to the transition matrix defined by~\eqtwo{red-Mf,red-KPL} works out as
\begin{align*}
  \mMf = 
  & \left[
    \ftrate{OP}\ftrate{PO}(\ftrate{PQ}+\ftrate{PU}-\ftrate{OU})
    -
    \frac{\dd{\ftrate{PO}}/\dd\at}{\trate{PO}+\trate{OP}}
    \right] 
  \\ 
  & 
    \times
    (\trate{PQ}+\trate{PU}-\trate{OU} )
    \,\ve1\ve1\T 
  \\ 
  &
    -(\trate{PQ}+\trate{PU}-\trate{OU})\ftrate{PO}\ftrate{QP}
    \,\ve1\ve2\T
  \\
  &
    -(\trate{PQ}+\trate{PU}-\trate{OU} )(\ftrate{PO}\ftrate{UP}-
    \ftrate{OP}\ftrate{UO})
    \,\ve1\ve6\T
  \\
  &
    + \left[
    \frac{\dd\ftrate{PO}/\dd\at}{\trate{PO}+\trate{OP}}
    -
    \ftrate{OP}\ftrate{PO}(\ftrate{PQ}+\ftrate{PU}-\ftrate{OU})
    \right]
  \\
  & \times
    \trate{PQ} \, 
    \,\ve2\ve1\T
  \\
  &
  + \trate{PQ}\ftrate{PO}\ftrate{QP}\,\ve2\ve2\T
  \\
  &
    + \trate{PQ}(\ftrate{PO}\ftrate{UP}- \ftrate{OP}\ftrate{UO})\,\ve2\ve6\T
  \\
  &
    + \left[
    \frac{\dd\ftrate{PO}/\dd\at}{\trate{PO} + \trate{OP}}
    - 
    \ftrate{OP}\ftrate{PO}(\ftrate{PQ}+\ftrate{PU}-\ftrate{OU})
    \right]
  \\
  & \times
    (\trate{PU} -\trate{OU} )
    \,\ve6\ve1\T
  \\
  &
    + (\trate{PU} -\trate{OU})\ftrate{PO}\ftrate{QP}\,\ve6\ve2\T
  \\
  &
    +
   (\trate{PU} -\trate{OU} )(\ftrate{PO}\ftrate{UP}-
   \ftrate{OP}\ftrate{UO}) \,\ve6\ve6\T .
\end{align*}
The Markov chain of the $\ina$ channel is linked to the rest of the cell
excitability model via the state $\xO$ which is the probability of the channel
being open, so we need to compute $\xO$ in terms of the new dynamic
variables. This is obtained from
\[
  \vx = \vUx +\eps \vvx,
\]
where $\vUx$ is given by~\eq{projection-leading} and $\vvx$ is given
by~\eq{vvx-OP}, with \eq{sol-across-ina} giving $\B_9$. This leads to 
\begin{align} \eqlabel{xO-OP}
  \xO = \frac{\trate{PO}}{\trate{PO} + \trate{OP}}\aN - \eps \B_9,
\end{align}
where
\begin{align} \eqlabel{B9-OP}
\B_9 = &
  \left(\trate{PO}+\trate{OP}\right)^{-1} \dfat{\ftrate{PO}}\aN
  \\ \nonumber &
  -\ftrate{OP}\,\ftrate{PO}(\ftrate{PQ}+\ftrate{PU}-\ftrate{OU})\aN  
  \\ \nonumber &
  + \ftrate{PO}\,\ftrate{QP}\aQ
  +(\ftrate{PO}\,\ftrate{UP}- \ftrate{OP}\,\ftrate{UO})\aU .
\end{align}

Matrix~$\mMf$ and coordinate $\B_9$ depend on time derivatives of the transition rates, which in
fact depend on the transmembrane voltage, hence the time derivative are
to be calculated  by the chain rule, \eg\ 
%
% \begin{align*}
%   \df{\left(\trate{OP}(\vm(\at))\right)}{\at} =& \df{\left(\trate{OP}(\vm(\at))\right)}{\vm} \ddt{\vm}, \\
%   \df{\left(\trate{PO}(\vm(\at))\right)}{\at} =& \df{\left(\trate{PO}(\vm(\at))\right)}{\vm} \ddt{\vm}.
% \end{align*}
%
\begin{align*}
  \dfat{}\left(\trate{OP}\right) =& \df{}{\vm}\left(\trate{OP}\right)\,\ddt{\vm}, \\
  \dfat{}\left(\trate{PO}\right) =& \df{}{\vm}\left(\trate{PO}\right)\,\ddt{\vm}.
\end{align*}

\begin{figure*}
  \centering
  \includegraphics{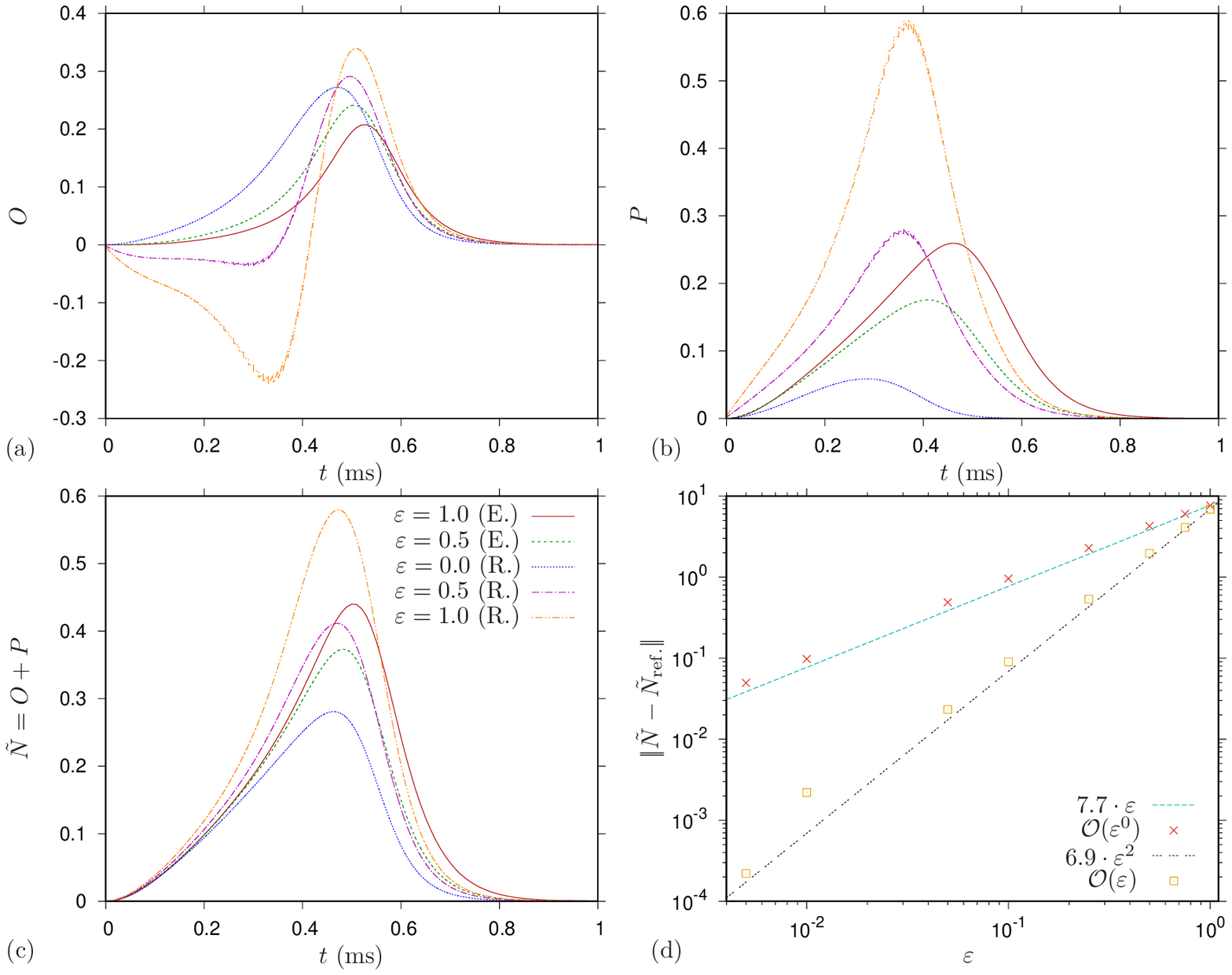}
    \caption{(a-c) Evolution of state occupancy in $OP$-embedded (E.) and
      $OP$-reduced (R.) model, and (d) error analysis of $OP$-reduced
      model. State $\xO$ occupancy (a), state $\xP$ occupancy (b),
      and state $\aN$ occupancy (c). The key in (c) applied to plots (a-c):
      the original model is denoted by red lines, the $OP$-embedded
      model $\eps = 0.5$ is shown with green lines, the reduced model
      without correction term ($\eps = 0$) is shown with blue lines,
      the $OP$-reduced model with correction term for $\eps = 0.5$ is
      shown with magenta lines, and the $OP$-reduced model with
      correction term for $\eps = 1.0$ is shown with orange lines.
      Panel (D) shows the order of approximation in $\eps$ for the
      leading-order reduced model (red crosses), first-order reduced
      model (yellow squares).  The norms were computed as a difference
      between the simulations of $\aN$ at time step $\delt=0.01\,\ms$
      and simulations of $\aNref=\xO+\xP$ computed with a
      time step of $\delt=5\cdot10^{-5}$~ms in the original model
      using the same value of $\eps$.  
      The cyan and grey straight lines are best fits
        by the corresponding powers of $\eps$. The data are
      shown on double logarithmic scale.  }
  \label{fig:reduction-OP}
\end{figure*}

Figure~\ref{fig:reduction-OP} shows the simulation results in the
$\xO\xP$-embeddings and corresponding $\xO\xP$-reduction.  The results should
be compared against the original model shown by red lines.
The simulations were done using extracted Markov chain model of $\ina$
driven by recordings of membrane voltage from whole cell simulations saved
every $0.01\,\ms$ and interpolated as necessary.
The state $\xO$ in reduced model was computed using 
\eqtwo{xO-OP,B9-OP}. 
For comparison of the reduced model with the embedding, 
the occupancy of state $\aN$ in the
embedded model was found as $\aN = \xO+\xP$.  

The simulations with leading order approximation (blue lines) show relatively
large deviation from the original model. The first-order accurate asymptotic
model computed for $\eps = 0.5$ (magenta lines) provides
better approximation than only the leading order term, 
however the state $\xO$ in this approximation goes below zero, which does not
make sense physically, as it represents a probability, so should be in the
interval $[0,1]$: note that the generic asymptotic theory does not
take into account these specifics.

The panel (d) shows the error norms computed using the following formula
\begin{align}
  \norm{\aN - \aNref} = \left[
  \int_0^\stmax (\aN(\st) - \aNref(\st))^2 \, \dd \st
  \right]^{1/2}
\end{align}
where $\aNref$ is the reference solution obtained for a very small time step,
and comparison is done for the interval of $\stmax=2$~ms of time-evolution.
The error norms
increase monotonically with $\eps$ and show the
convergence for the leading-order and first-order
approximations as expected, which confirms the correctness of the formulas.

\subsection{STU-embedding and reduction of S, T and U into M}

In this section we develop another approximation of the original system, which
considers the transitions between states $\xS$, $\xT$ and $\xU$ as fast, which
in asymptotics leads to their merger into a new state $\aM$.  This choice is
supported by the empirical embedding procedure as described
in~\secn{embedding}, details can be found in~\refcite{Stary-2016}. Now the
matrix of fast transition rates is
\begin{align} \eqlabel{dINa-STU-A0}
      \mAf = & \Mx{
      0&0&0&0& 0          & 0          & 0          &0&0\\
      0&0&0&0& 0          & 0          & 0          &0&0\\
      0&0&0&0& 0          & 0          & 0          &0&0\\
      0&0&0&0& 0          & 0          & 0          &0&0\\
      0&0&0&0& \DF        & \trate{TS} & 0          &0&0\\
      0&0&0&0& \trate{ST} & \DF        & \trate{UT} &0&0\\      
      0&0&0&0& 0          & \trate{TU} & \DF        &0&0\\
      0&0&0&0& 0          & 0          & 0          &0&0\\
      0&0&0&0& 0          & 0          & 0          &0&0
    } 
\end{align}
and the remaining, slow rates constitute the matrix
\begin{align} \eqlabel{dINa-STU-A1}
  \mAs = & \Mx{
\DF    &\TR{PO}&0      &0      &0      &0      &\TR{UO}&0      &0      \\
\TR{OP}&\DF    &\TR{QP}&0      &0      &0      &\TR{UP}&0      &0      \\
0      &\TR{PQ}&\DF    &\TR{RQ}&0      &\TR{TQ}&0      &0      &0      \\
0      &0      &\TR{QR}&\DF    &\TR{SR}&0      &0      &0      &0      \\
0      &0      &0      &\TR{RS}&\DF    &0      &0      &0      &0      \\
0      &0      &\TR{QT}&0      &0      &\DF    &0      &0      &0      \\
\TR{OU}&\TR{PU}&0      &0      &0      &0      &\DF    &\TR{VU}&0      \\
0      &0      &0      &0      &0      &0      &\TR{UV}&\DF    &\TR{WV}\\
0      &0      &0      &0      &0      &0      &0      &\TR{VW}&\DF
}
\end{align}
The right eigenvectors corresponding to zero eigenvalue of this system 
can be chosen as 
\begin{align} \eqlabel{STU-evects}
  \vevecA\iin &= \ve\iin, \qquad \iin=1,\dots,4, \\\nonumber
  \vevecA5 &= (\trate{UT}\trate{TS}+
              \trate{UT}\trate{ST}+\trate{TU}\trate{ST})^{-1}  \\\nonumber
            & \times\left(\trate{UT}\trate{TS}\,\ve5
              +\trate{UT}\trate{ST}\,\ve6
              +\trate{TU}\trate{ST}\,\ve7\right),\\\nonumber
  \vevecA\iin &= \ve{\iin+2}, \qquad \iin=6,7.
\end{align}
% \begin{align} \eqlabel{STU-evects}
%   \vevecA1 =& \Mx{1,&0,&0,&0,&0,&0,&0,&0,&0}\T , \\\nonumber
%   \vevecA2 =& \Mx{0,&1,&0,&0,&0,&0,&0,&0,&0}\T , \\\nonumber
%   \vevecA3 =& \Mx{0,&0,&1,&0,&0,&0,&0,&0,&0}\T , \\\nonumber
%   \vevecA4 =& \Mx{0,&0,&0,&1,&0,&0,&0,&0,&0}\T , \\\nonumber
%   \vevecA5 =& (\trate{UT}\trate{TS}+ \trate{UT}\trate{ST}+\trate{TU}\trate{ST})^{-1}  \\\nonumber
%             & \times\Mx{0,&0,&0,&0,&\trate{UT}\trate{TS},&
%               \trate{UT}\trate{ST},&\trate{TU}\trate{ST},&0,&0}\T ,\\\nonumber
%   \vevecA6 =& \Mx{0,&0,&0,&0,&0,&0,&0,&1,&0}\T , \\\nonumber
%   \vevecA7 =& \Mx{0,&0,&0,&0,&0,&0,&0,&0,&1}\T.
% \end{align}
%
The corresponding left eigenvectors are
\begin{align} \eqlabel{STU-eadvects}
  \vadvecA\iin = & \vevecA\iin, \qquad \iin=1,2,3,4,6,7, \nonumber\\
  \vadvecA{5}  = & \ve5+\ve6+\ve7. 
\end{align}
With account of these, we can keep the names of the original dynamic variables
for all states except $\xS$, $\xT$, $\xU$, so the vector of states of the
reduced system is
\begin{align}
  \va = [\xO, \xP,\xQ,\xR,\aM,\xV,\xW]\T .
\end{align}

These are all the ingredients needed for the derivation of the leading-order
approximation. We have $\vadvecA\iin'=0$ for all $\iin=1,\dots,7$
so~\eq{red-Ml} gives the leading-order transition rate matrix for the reduced
model as
\begin{align} \eqlabel{STU-M0}
  \mMl = \Mx{
\DF       &\trate{PO}&0         &0         &\trate{MO}&0         &0         \\
\trate{OP}&\DF       &\trate{QP}&0         &\trate{MP}&0         &0         \\
0         &\trate{PQ}&\DF       &\trate{RQ}&\trate{MQ}&0         &0         \\
0         &0         &\trate{QR}&\DF       &\trate{MR}&0         &0         \\
\trate{OM}&\trate{PM}&\trate{QM}&\trate{RM}&\DF       &\trate{VM}&0         \\
0         &0         &0         &0         &\trate{MV}&\DF       &\trate{WV}\\
0         &0         &0         &0         &0         &\trate{VW}&\DF
  },
\end{align}
with the new transition rates defined as
\begin{align}
  \trate{MO} =&  \trate{UO}\,\STUtrate{STTU}, \hspace{2em} \trate{OM} =\trate{OU}, \\ \nonumber
  \trate{MP} =&  \trate{UP}\,\STUtrate{STTU}, \hspace{2em} \trate{PM} =\trate{PU}, \\\nonumber
  \trate{MQ} =&  \trate{TQ}\,\STUtrate{UTST}, \hspace{2em} \trate{QM} =\trate{QT},\\\nonumber
  \trate{MR} =&  \trate{SR}\,\STUtrate{UTTS}, \hspace{2em} \trate{RM} =\trate{RS},\\ \nonumber
  \trate{MV} =&  \trate{UV}\,\STUtrate{STTU}, \hspace{2em} \trate{VM} =\trate{VU}.\\\nonumber
\end{align}
These expression use the notation $\STUtrate{\mathrm{ijkl}}$ as an
abbreviation for
\begin{align}
  \STUtrate{\mathrm{ijkl}} = \frac{
    \trate{\mathrm{ij}}\,\trate{\mathrm{kl}}
  }{
  \trate{UT}\trate{TS}+ \trate{UT}\trate{ST}+ \trate{ST}\trate{TU}
  }.
\end{align}

The original coordinates are recovered from the reduced one by 
\begin{subequations}
  \begin{align}
    \xS =& \STUtrate{UTTS} \aM, \\
    \xT =& \STUtrate{UTST} \aM, \\
    \xU =& \STUtrate{STTU} \aM.
  \end{align}
\end{subequations}

As can be seen in~\fig{ap-red} below, the quality of the approximation
obtained with these asymptotics, is very good. This was of course to be
expected based on the results of the empirical embedding study, as discussed
above.

\subsection{Embedding and reduction of R and Q states of STU-reduction into L}

In this section, we investigate how one can build on the success of the
$\xS\xT\xU$ embedding and achieve further reduction. As we have already
considered the $\xO\xP$ reduction above, we now consider $\xR\xQ$
reduction. That is, we consider the transition rates between $\xR$ and $\xQ$
as fast, which will lead to the merger of these two states into a new state
$\aL$. So in the context of the present section, the ``original model'' is
defined by the matrix~\eq{STU-M0}, which will now be called $\mM$, 
while the fast matrix in the new embedding is 
\begin{align}
  \mMfa = \Mx{
    0 & 0 & 0          & 0          & 0 & 0 & 0 \\
    0 & 0 & 0          & 0          & 0 & 0 & 0 \\
    0 & 0 & \DF        & \trate{RQ} & 0 & 0 & 0 \\
    0 & 0 & \trate{QR} & \DF        & 0 & 0 & 0 \\
    0 & 0 & 0          & 0          & 0 & 0 & 0 \\
    0 & 0 & 0          & 0          & 0 & 0 & 0 \\
    0 & 0 & 0          & 0          & 0 & 0 & 0 
  } , 
\end{align}
and the slow matrix is 
\begin{align}
  \mMs = \Mx{
    \DF       &\trate{PO}&0         &0         &\trate{MO}&0         &0         \\
    \trate{OP}&\DF       &\trate{QP}&0         &\trate{MP}&0         &0         \\
    0         &\trate{PQ}&\DF       &0         &\trate{MQ}&0         &0         \\
    0         &0         &0         &\DF       &\trate{MR}&0         &0         \\
    \trate{OM}&\trate{PM}&\trate{QM}&\trate{RM}&\DF       &\trate{VM}&0         \\
    0         &0         &0         &0         &\trate{MV}&\DF       &\trate{WV}\\
    0         &0         &0         &0         &0         &\trate{VW}&\DF
  }
\end{align}
Acting as before, we find the right eigenvectors of $\mMfa$ corresponding to
zero eigenvalue as
\begin{align*}
  \vevecA\iin & = \ve\iin, \qquad \iin=1,2,\\
  \vevecA{3} & =(\trate{QR} + \trate{RQ})^{-1} \left(\trate{QR}\,\ve3+\trate{RQ}\,\ve4\right),\\
  \vevecA\iin & = \ve{\iin+1}, \qquad \iin=4,5,6,
\end{align*}
%
% \begin{align*}
%   \vevecA{1} = &\Mx{1,&0,&0,&0,&0,&0,&0}\T,\\
%   \vevecA{2} = &\Mx{0,&1,&0,&0,&0,&0,&0}\T,\\
%   \vevecA{3} = &(\trate{QR} + \trate{RQ})^{-1}
%                 \Mx{0,&0,&\trate{QR},&\trate{RQ},&0,&0,&0}\T,\\
%   \vevecA{4} = &\Mx{0,&0,&0,&0,&1,&0,&0}\T,\\
%   \vevecA{5} = &\Mx{0,&0,&0,&0,&0,&1,&0}\T,\\
%   \vevecA{6} = &\Mx{0,&0,&0,&0,&0,&0,&1}\T,
% \end{align*}
%
the corresponding left eigenvectors  as
\begin{align*}
  \vadvecA\iin = & \vevecA{\iin}, \qquad \iin = 1, 2, 4, 5, 6, \\
  \vadvecA{3} = & \ve3+\ve4,
\end{align*}
% \begin{align*}
%   \vadvecA\iin = & \vevecA{\iin}, \qquad \iin = 1, 2, 4, 5, 6, \\
%   \vadvecA{3} = & \Mx{0,&0,&1,&1,&0,&0,&0}\T,
% \end{align*}
%
and we set the names of the components of the reduced vector as
\begin{align*}
  \va = [\xO, \xP,\aL,\aM,\xV,\xW]\T .
\end{align*}

The resulting leading-order reduced matrix works out as
\begin{align}
  \mMl = \Mx{
    \DF       &\trate{PO}&0         &\trate{MO}&0         &0         \\
    \trate{OP}&\DF       &\trate{LP}&\trate{MP}&0         &0         \\
    0         &\trate{PL}&\DF       &\trate{ML}&0         &0         \\
    \trate{OM}&\trate{PM}&\trate{LM}&\DF       &\trate{VM}&0         \\
    0         &0         &0         &\trate{MV}&\DF       &\trate{WV}\\
    0         &0         &0         &0         &\trate{VW}&\DF
  }
\end{align}
with the new transition rates defined by
\begin{align} \eqlabel{STUQR-M0}
  \trate{LP} =& \trate{QP}\QRtrate{QR}, && \trate{PL} = \trate{PQ}\\ \nonumber              
  \trate{LM} =& \trate{QM}\QRtrate{QR} + \trate{RM}\QRtrate{RQ}, && \trate{ML} = \trate{MQ} + \trate{MR},
\end{align}
where
\begin{align}
  \QRtrate{\mathrm{ij}} = 
  \frac{\trate{\mathrm{ij}}}{\trate{QR} + \trate{RQ}}.
\end{align}
It is easily seen that the resulting reduced
  model~\eq{STUQR-M0} will be the same if, instead, we do the $QR$
  reduction first and $STU$ reduction second, or do them
  simultaneously, \ie\ include all of $\TT ST$, $\TT TU$ and $\TT QR$
  in $\mAf$ in the first place.

\begin{figure*}
  \centering
  \includegraphics{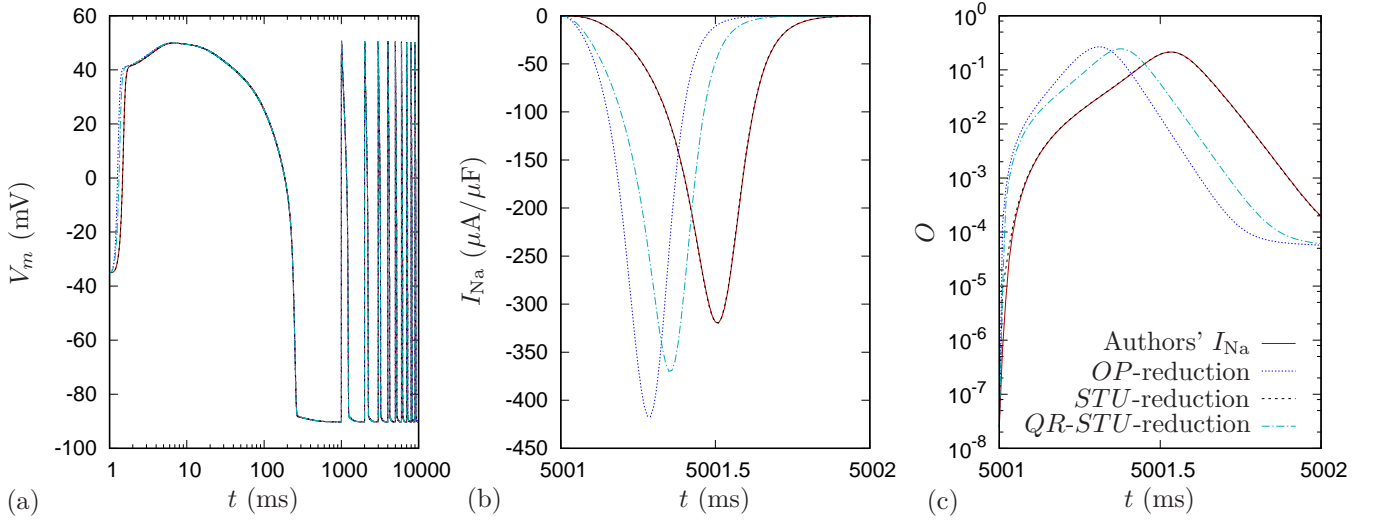}
  \caption{Comparison of action potential solutions produced with
    different asymptotics of the $\ina$ channel model. %
    (a) Transmembrane voltage during 10 action potential vs time (log scale). %
    (b) The $\ina$ current at the onset of the fifth action potential. %
    (c) The state occupancy of $\xO$ (log scale) during the fifth action potential. %
    Red solid line: the original model. %
    Blue dotted line: the $\xO\xP$-reduction. %
    Black dashed line: the $\xS\xT\xU$-reduction. %
    Cyan dash-dotted line: $\xQ\xR$-$\xS\xT\xU$ reduction. }
  \label{fig:ap-red}
\end{figure*}

\Fig{ap-red} presents the results of the $\xQ\xR$-$\xS\xT\xU$,
together with
the previously considered $\xO\xP$ and $\xS\xT\xU$ reductions. 
In these simulations, the $\ina$ channel model was not
  driven by the recorded $\vm(\st)$ as before, but rather was part of the
  full cell model \eq{fullmodel}.
The full original model and the three reduced version
were run in the same protocol, which included stimulation with a period of one
second, starting from $\st=1\,\ms$ (this was done in order to be able to show
the time in panel (a) in the logarithmic scale). One can see that the reduced
models are indistinguishable from the full model except for the upstroke of
the action potential. The upstroke of the fifth action potential is shown in
detail in panels (b) and (c), for the probability of the $\ina$ channel being
open, and the resulting value of this current. We see that the results
generally agree with what could be expected from the empirical embedding
studies illustrated in~\fig{embeddings}. Namely, the $\xO\xP$ embedding gives
a rather poor approximation, the $\xQ\xR$-$\xS\xT\xU$ embedding is slightly
better, while $\xS\xT\xU$ is very good. 

\section{Discussion}
\seclabel{discussion}

\begin{figure}
  \centering
  \includegraphics[width=\columnwidth]{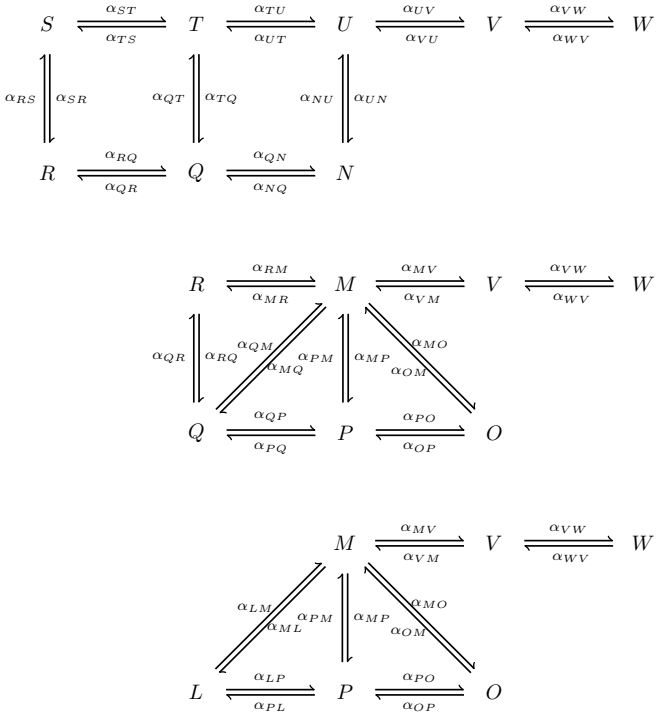}
  \caption{Diagram of reduced Markov chain models of $\ina$
    channel. Top diagram shows the $\xO\xP$-reduced model, middle diagram
    shows $STU$-reduced model, and bottom diagram shows
    $RQ$-$STU$-reduced model.}
  \label{fig:diag-tr-1}
\end{figure}

\Fig{diag-tr-1} summarises the Markov chain models occurring as a result
of the three asymptotics we have considered: this is to be compared with the
original scheme shown in~\fig{diag-tr}. 

Asymptotic reduction based on time scale separation can pursue at least two
different goals: reducing the number of dynamic equations, and reducing
stiffness of those equations. The reductions considered in this paper are not
particularly impressive in terms of reducing the number of equations: we have
reduced by maximum of three out of nine, which is even less significant in
comparison with the number of other dynamic equations in a typical model of an
excitable cell, beyond the Markov chain of the $\ina$
channel. However, 
in practical applications the main goal is the other one: reducing the stiffness. 
To achieve a simple practical estimate of this characteristic, we measured the
stiffness of the model by the maximum time step size $\dt$ which provides a
stable solution using the forward Euler solver for the isolated $\ina$ model
driven by a recorded action potential. The original full model allows the time
step of about $\dt\approx0.04\,\ms$ for stable computations; an increase above
that leads to numerical instability. In comparison to that, all three models
considered allow $\dt\approx0.044\,\ms$, \ie\ a rather modest improvement.
The limited progress in this is due to the fact that in all three examples
considered, we have included in the embedding only some of the fastest
transition rates. And even in these cases, we have seen that asymptotic
removal of some of the fast processes affects the accuracy of
computations. Even though these effects are seen only during the
upstrokes of the action potential, these upstrokes are of principal
significance as they determine the conduction velocity in
spatially-distributed simulation, and therefore also the more delicate and
more important phenomena such as conduction block, wavebreaks \etc. Hence
further increase of the number of the reduced degrees of freedom does not seem
to be an answer. Further research is of course needed to establish that with
  certainty, but, as already noted above, \eg\ the ``straightforward''
  approach embedding \emph{all} the transition rates that appear ``fast'' in
  \fig{diag-tr} does not yield a satisfactory
  approximation~\cite{Stary-2016}.

\begin{figure}
  \centering
  \includegraphics{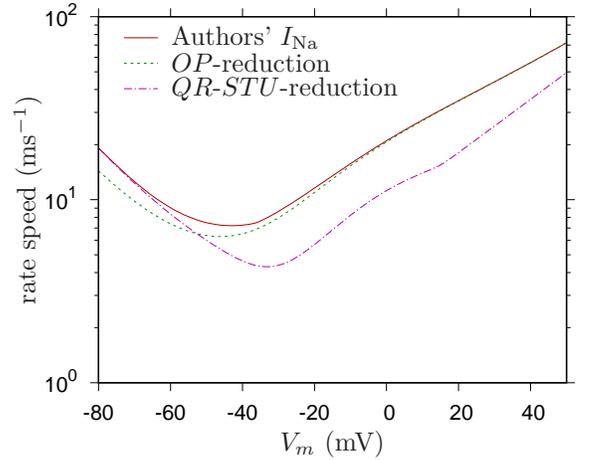}
  \caption{ Largest absolute values of the eigenvalues of the transition
    matrices for the original Markov chain and the two selected reduced
    models.}
  \label{fig:stiff}
\end{figure}

From a more theoretical viewpoint, stiffness can
  sometimes be characterized by the eigenvalues of the system; in
  particular, the upper limit of the integration step is mainly
  affected by the eigenvalue with the largest absolute value.
  In~\fig{stiff} we plot those absolute values for the original model 
  and the reduced models, as functions of the transmembrane
  voltage. We see that whereas $OP$ reduction somewhat reduces
  stiffness at the lower end of the $\vm$ scale, it has virtually no
  effect at the upper end. On the contrary, $QR$-$STU$ reduction
  noticeably reduces stiffness at the upper end, without changing it
  at the lower end.

        So, as far as the question posed in the introduction is
        concerned, the results obtained here seem to suggest that
        Tikhonov asymptotic structure, which implies fixed
        distribution of the roles of ``fast'' and ``slow'' variables,
        or, in this particular class of applications, rather ``fast''
        and ``slow'' transition rates, may not be quite adequate for
        the this particular model of the fast sodium current, and some
        non-Tikhonov parametric embedding may be more fruitful, say
        when transition rates are considered fast in one range of
        $\vm$ and slow in the complementary range, possibly with the
        asymmetry between the reciprocal rates taken into
        account.

An alternative approach, which has proved to be more practical than the one
considered here, has been described in our previous
works~\cite{Stary2015,Stary-Biktashev-2015-CinC}, dubbed ``exponential
solvers''. However, that approach is
purely numerical and does not explicitly take into account the fast-slow
structure of the model, hence an asymptotic approach seems to have an
\apriori\ advantage, which ought to have been explored. We hope that the
present study fills this gap to a certain extent. 

An attractive possibility to improve the accuracy of the asymptotics and hence
to open the way to further decrease the number of equations and reduce the
stiffness, seems to be using higher-order asymptotics. We have explored this
only in one of the three examples, but it already shows that (i) the algebraic
complexity of the resulting formulas increases considerably, (ii) more
significantly, some improvement in accuracy is devalued by the fact that the
resulting model, unlike the leading-order asymptotics, no longer behaves as a
``proper'' Markov chain: the vector of dynamic variables is not guaranteed to
remain stochastic, in particular, it can easily lead to negative values of the
state occupancies. This happens because we have used the asymptotic
theory which was designed for generic systems and is not tailored for the
specific requirements of Markov chains. Hence another possible way for improvement
may be in developing higher-order asymptotics strictly within the class of
Markov chains.

\begin{acknowledgments}
VNB gratefully acknowledges the current
financial support of the EPSRC via grant EP/N014391/1 (UK)
TS acknowledges financial support of the University of Exeter via PhD
Studentship and of the EPSRC via grant EP/N024508/1.
\end{acknowledgments}

\appendix

\section{Derivation of Reduced System}
\seclabel{reduct-comp}

We use the Taylor expansion for the functions $\vf(\vu)$ and
$\vh(\vu)$, such that, after substitution of the sought solution
\eqref{eq:ansatz-solution}, we get the expression on the right hand
side of \eqref{eq:ode-def} as
\begin{align}
  \label{eq:ode-rhs}
  \vf(\vU) + \eps\sum_\iinj \pf{\vf}{\uu_\iinj} \V_\iinj
  + \eps^2\sum_{\iini,\iinii}\ppf{\vf}{\uu_\iini}{\uu_\iinii} \V_\iini\V_\iinii \\ \nonumber
  + \eps\vh(\vU) + \eps^2\sum_\iini \pf{\vh}{\uu_\iini} \V_\iini+
  \cO{3}.
\end{align}
The first term $\vf(\vU)=0$ by assumption, and in the second
term we note that the derivatives constitute the Jacobian matrix 
and expand the $\vv$ according to
\eqref{eq:solution-perturbation}. Then the previous expression
\eqref{eq:ode-rhs} rewrites as
\begin{align}
  \eps\mF(\vU)  \sum_\imn \B_\imn \vevec\imn 
  + \eps^2\sum_{\iini,\iinii}\ppf{\vf}{\uu_\iini}{\uu_\iinii} \V_\iini\V_\iinii + \\ \nonumber
  \eps\vh(\vU) + \eps^2\sum_\iini \pf{\vh}{\uu_\iini} \V_\iini+
  \cO{3}. \nonumber
\end{align}
We substitute the sought solution also to the left hand side of
\eqref{eq:ode-def}.  We use the knowledge of eigenvectors
corresponding to zero eigenvalues from \eqref{eq:evec-tang} and expand
the perturbed term \eqref{eq:solution-perturbation}. Then using a
chain rule for the derivative of $\vU(\va)$ and $\vevec\imn(\va)$ we
get
\begin{align}
    \label{eq:ode-lhs}
  \sum_\iim\pf{\vU}{\A_\iim}\ddt{\A_\iim} + \eps\ddt{\vv}  =
  \sum_\iim\vevec\iim\ddt{\A_\iim} +  \\ \nonumber
   \eps\sum_\imn\left(\ddt{\B_\imn} \vevec\imn +\B_\imn\df{\vevec\imn }{\A_\iim }\ddt{\A_\iim } \right).
\end{align}
Combining the right-hand side given by \eqref{eq:ode-rhs} and the left-hand side
given by \eqref{eq:ode-lhs}, we rewrite \eqref{eq:ode-def} as
\begin{align} \label{eq:ode-subs-solution}
  & \sum_\iim\vevec\iim\ddt{\A_\iim} +
  \eps\sum_\imn\left(\ddt{\B_\imn} \vevec\imn +\B_\imn\df{\vevec\imn}{\A_\iim }\ddt{\A_\iim } \right) 
  \\ \nonumber &
   % = \vf(\vU + \eps \vv) + \eps\vh(\vU + \eps \vv) 
  = \eps\mF \sum_\imn \B_\imn \vevec\imn 
  + \eps^2\sum_{\iini,\iinii}\ppf{\vf}{\uu_\iini}{\uu_\iinii} \V_\iini\V_\iinii 
  \\ \nonumber &
  + \eps\vh(\vU) + \eps^2\sum_\iini \pf{\vh}{\uu_\iini} \V_\iini+
  \cO{3}. 
\end{align}
Multiplying the equation by the adjoint eigenvectors $\vadvecT{\iin}$
gives
\begin{align} \label{eq:ode-rewriten}
 \epsi\ddt{\A_\iin} +
  \ddt{\B_\iin}
  = &
  \eval\iin\B_\iin  +
  \vadvecT{\iin}\vh(\vU)  \\
    &  + \eps\vadvecT{\iin}\left[\sum_\iinj \pf{\vh}{\uu_\iinj} \V_\iinj+
  \sum_{\iini,\iinii}\ppf{\vf}{\uu_\iini}{\uu_\iinii}
  \V_\iini\V_\iinii 
  \right. \nonumber \\ & \left.
    - \epsi\sum_{\imn,\iim} \B_\imn\pf{\vevec\imn }{\A_\iim }\ddt{\A_\iim }\right]+
  \cO{2} .\nonumber
\end{align}
Considering separately the zero and the stable eigenvalues then yields
equations~\eq{split-ode-a} and \eq{split-ode-b} respectively. 

%\bibliography{amcm}

\begin{thebibliography}{26}%
\makeatletter
\providecommand \@ifxundefined [1]{%
 \@ifx{#1\undefined}
}%
\providecommand \@ifnum [1]{%
 \ifnum #1\expandafter \@firstoftwo
 \else \expandafter \@secondoftwo
 \fi
}%
\providecommand \@ifx [1]{%
 \ifx #1\expandafter \@firstoftwo
 \else \expandafter \@secondoftwo
 \fi
}%
\providecommand \natexlab [1]{#1}%
\providecommand \enquote  [1]{``#1''}%
\providecommand \bibnamefont  [1]{#1}%
\providecommand \bibfnamefont [1]{#1}%
\providecommand \citenamefont [1]{#1}%
\providecommand \href@noop [0]{\@secondoftwo}%
\providecommand \href [0]{\begingroup \@sanitize@url \@href}%
\providecommand \@href[1]{\@@startlink{#1}\@@href}%
\providecommand \@@href[1]{\endgroup#1\@@endlink}%
\providecommand \@sanitize@url [0]{\catcode `\\12\catcode `\$12\catcode
  `\&12\catcode `\#12\catcode `\^12\catcode `\_12\catcode `\%12\relax}%
\providecommand \@@startlink[1]{}%
\providecommand \@@endlink[0]{}%
\providecommand \url  [0]{\begingroup\@sanitize@url \@url }%
\providecommand \@url [1]{\endgroup\@href {#1}{\urlprefix }}%
\providecommand \urlprefix  [0]{URL }%
\providecommand \Eprint [0]{\href }%
\providecommand \doibase [0]{http://dx.doi.org/}%
\providecommand \selectlanguage [0]{\@gobble}%
\providecommand \bibinfo  [0]{\@secondoftwo}%
\providecommand \bibfield  [0]{\@secondoftwo}%
\providecommand \translation [1]{[#1]}%
\providecommand \BibitemOpen [0]{}%
\providecommand \bibitemStop [0]{}%
\providecommand \bibitemNoStop [0]{.\EOS\space}%
\providecommand \EOS [0]{\spacefactor3000\relax}%
\providecommand \BibitemShut  [1]{\csname bibitem#1\endcsname}%
\let\auto@bib@innerbib\@empty
%</preamble>
\bibitem [{\citenamefont {Plank}\ \emph {et~al.}(2008)\citenamefont {Plank},
  \citenamefont {Zhou}, \citenamefont {Greenstein}, \citenamefont {Cortassa},
  \citenamefont {Winslow}, \citenamefont {O'Rourke},\ and\ \citenamefont
  {Trayanova}}]{Plank-etal-2008}%
  \BibitemOpen
  \bibfield  {author} {\bibinfo {author} {\bibfnamefont {G.}~\bibnamefont
  {Plank}}, \bibinfo {author} {\bibfnamefont {L.}~\bibnamefont {Zhou}},
  \bibinfo {author} {\bibfnamefont {J.~L.}\ \bibnamefont {Greenstein}},
  \bibinfo {author} {\bibfnamefont {S.}~\bibnamefont {Cortassa}}, \bibinfo
  {author} {\bibfnamefont {R.~L.}\ \bibnamefont {Winslow}}, \bibinfo {author}
  {\bibfnamefont {B.}~\bibnamefont {O'Rourke}}, \ and\ \bibinfo {author}
  {\bibfnamefont {N.~A.}\ \bibnamefont {Trayanova}},\ }\bibfield  {title}
  {\enquote {\bibinfo {title} {From mitochondrial ion channels to arrhythmias
  in the heart: computational techniques to bridge the spatio-temporal
  scales},}\ }\href@noop {} {\bibfield  {journal} {\bibinfo  {journal} {Philos
  Transact Roy Soc A}\ }\textbf {\bibinfo {volume} {366}},\ \bibinfo {pages}
  {3381--3409} (\bibinfo {year} {2008})}\BibitemShut {NoStop}%
\bibitem [{\citenamefont {Richards}\ \emph {et~al.}(2013)\citenamefont
  {Richards}, \citenamefont {Glosli}, \citenamefont {Draeger}, \citenamefont
  {Mirin}, \citenamefont {Chan}, \citenamefont {Fattebert}, \citenamefont
  {Krauss}, \citenamefont {Oppelstrup}, \citenamefont {Butler}, \citenamefont
  {Gunnels}, \citenamefont {Gurev}, \citenamefont {Kim}, \citenamefont
  {Magerlein}, \citenamefont {Reumann}, \citenamefont {Wen},\ and\
  \citenamefont {Rice}}]{Richards-etal-2013}%
  \BibitemOpen
  \bibfield  {author} {\bibinfo {author} {\bibfnamefont {D.~F.}\ \bibnamefont
  {Richards}}, \bibinfo {author} {\bibfnamefont {J.~N.}\ \bibnamefont
  {Glosli}}, \bibinfo {author} {\bibfnamefont {E.~W.}\ \bibnamefont {Draeger}},
  \bibinfo {author} {\bibfnamefont {A.~A.}\ \bibnamefont {Mirin}}, \bibinfo
  {author} {\bibfnamefont {B.}~\bibnamefont {Chan}}, \bibinfo {author}
  {\bibfnamefont {J.}~\bibnamefont {Fattebert}}, \bibinfo {author}
  {\bibfnamefont {W.~D.}\ \bibnamefont {Krauss}}, \bibinfo {author}
  {\bibfnamefont {T.}~\bibnamefont {Oppelstrup}}, \bibinfo {author}
  {\bibfnamefont {C.~J.}\ \bibnamefont {Butler}}, \bibinfo {author}
  {\bibfnamefont {J.~A.}\ \bibnamefont {Gunnels}}, \bibinfo {author}
  {\bibfnamefont {V.}~\bibnamefont {Gurev}}, \bibinfo {author} {\bibfnamefont
  {C.}~\bibnamefont {Kim}}, \bibinfo {author} {\bibfnamefont {J.}~\bibnamefont
  {Magerlein}}, \bibinfo {author} {\bibfnamefont {M.}~\bibnamefont {Reumann}},
  \bibinfo {author} {\bibfnamefont {H.~F.}\ \bibnamefont {Wen}}, \ and\
  \bibinfo {author} {\bibfnamefont {J.~J.}\ \bibnamefont {Rice}},\ }\bibfield
  {title} {\enquote {\bibinfo {title} {Towards real-time simulation of cardiac
  electrophysiology in a human heart at high resolution},}\ }\href@noop {}
  {\bibfield  {journal} {\bibinfo  {journal} {Computer Methods in Biomechanics
  and Biomedical Engineering}\ }\textbf {\bibinfo {volume} {16}},\ \bibinfo
  {pages} {802--805} (\bibinfo {year} {2013})}\BibitemShut {NoStop}%
\bibitem [{\citenamefont {Bondarenko}(2014)}]{Bondarenko-2014}%
  \BibitemOpen
  \bibfield  {author} {\bibinfo {author} {\bibfnamefont {V.~E.}\ \bibnamefont
  {Bondarenko}},\ }\bibfield  {title} {\enquote {\bibinfo {title} {A
  compartmentalized mathematical model of the {$\beta_1$}-adrenergic signaling
  system in mouse ventricular myocytes},}\ }\href@noop {} {\bibfield  {journal}
  {\bibinfo  {journal} {PLOS ONE}\ }\textbf {\bibinfo {volume} {9}},\ \bibinfo
  {pages} {e89113} (\bibinfo {year} {2014})}\BibitemShut {NoStop}%
\bibitem [{\citenamefont {Hodgkin}\ and\ \citenamefont
  {Huxley}(1952)}]{Hodgkin-Huxley-1952}%
  \BibitemOpen
  \bibfield  {author} {\bibinfo {author} {\bibfnamefont {A.~L.}\ \bibnamefont
  {Hodgkin}}\ and\ \bibinfo {author} {\bibfnamefont {A.~F.}\ \bibnamefont
  {Huxley}},\ }\bibfield  {title} {\enquote {\bibinfo {title} {A quantitative
  description of membrane current and its application to conduction and
  excitation in nerve},}\ }\href@noop {} {\bibfield  {journal} {\bibinfo
  {journal} {J Physiol Lond}\ }\textbf {\bibinfo {volume} {117}},\ \bibinfo
  {pages} {500--544} (\bibinfo {year} {1952})}\BibitemShut {NoStop}%
\bibitem [{\citenamefont {Rush}\ and\ \citenamefont
  {Larsen}(1978)}]{Rush-Larsen-1978}%
  \BibitemOpen
  \bibfield  {author} {\bibinfo {author} {\bibfnamefont {S.}~\bibnamefont
  {Rush}}\ and\ \bibinfo {author} {\bibfnamefont {H.}~\bibnamefont {Larsen}},\
  }\bibfield  {title} {\enquote {\bibinfo {title} {A practical algorithm for
  solving dynamic membrane equations},}\ }\href@noop {} {\bibfield  {journal}
  {\bibinfo  {journal} {IEEE Trans BME}\ }\textbf {\bibinfo {volume} {25}},\
  \bibinfo {pages} {389--392} (\bibinfo {year} {1978})}\BibitemShut {NoStop}%
\bibitem [{\citenamefont {Perego}\ and\ \citenamefont
  {Veneziani}(2009)}]{Perego-Veneziani-2009}%
  \BibitemOpen
  \bibfield  {author} {\bibinfo {author} {\bibfnamefont {M.}~\bibnamefont
  {Perego}}\ and\ \bibinfo {author} {\bibfnamefont {A.}~\bibnamefont
  {Veneziani}},\ }\bibfield  {title} {\enquote {\bibinfo {title} {An efficient
  generalization of the {Rush}-{Larsen} method for solving electro-physiology
  membrane equations},}\ }\href@noop {} {\bibfield  {journal} {\bibinfo
  {journal} {{Electronic Transactions on Numerical Analysis}}\ }\textbf
  {\bibinfo {volume} {35}},\ \bibinfo {pages} {234--256} (\bibinfo {year}
  {2009})}\BibitemShut {NoStop}%
\bibitem [{\citenamefont {Marsh}, \citenamefont {Ziaratgahi},\ and\
  \citenamefont {Spiteri}(2012)}]{Marsh-etal-2012}%
  \BibitemOpen
  \bibfield  {author} {\bibinfo {author} {\bibfnamefont {M.~E.}\ \bibnamefont
  {Marsh}}, \bibinfo {author} {\bibfnamefont {S.~T.}\ \bibnamefont
  {Ziaratgahi}}, \ and\ \bibinfo {author} {\bibfnamefont {R.~J.}\ \bibnamefont
  {Spiteri}},\ }\bibfield  {title} {\enquote {\bibinfo {title} {The secrets to
  the success of the {Rush}-{Larsen} method and its generalizations},}\
  }\href@noop {} {\bibfield  {journal} {\bibinfo  {journal} {IEEE Trans BME}\
  }\textbf {\bibinfo {volume} {59}},\ \bibinfo {pages} {2506--2515} (\bibinfo
  {year} {2012})}\BibitemShut {NoStop}%
\bibitem [{\citenamefont {Stary}\ and\ \citenamefont
  {Biktashev}(2015{\natexlab{a}})}]{Stary2015}%
  \BibitemOpen
  \bibfield  {author} {\bibinfo {author} {\bibfnamefont {T.}~\bibnamefont
  {Stary}}\ and\ \bibinfo {author} {\bibfnamefont {V.~N.}\ \bibnamefont
  {Biktashev}},\ }\bibfield  {title} {\enquote {\bibinfo {title} {Exponential
  integrators for a {Markov} chain model of the fast sodium channel of
  cardiomyocytes.}}\ }\href {\doibase 10.1109/TBME.2014.2366466} {\bibfield
  {journal} {\bibinfo  {journal} {IEEE Trans BME}\ }\textbf {\bibinfo {volume}
  {62}},\ \bibinfo {pages} {1070--1076} (\bibinfo {year}
  {2015}{\natexlab{a}})}\BibitemShut {NoStop}%
\bibitem [{\citenamefont {Stary}\ and\ \citenamefont
  {Biktashev}(2015{\natexlab{b}})}]{Stary-Biktashev-2015-CinC}%
  \BibitemOpen
  \bibfield  {author} {\bibinfo {author} {\bibfnamefont {T.}~\bibnamefont
  {Stary}}\ and\ \bibinfo {author} {\bibfnamefont {V.}~\bibnamefont
  {Biktashev}},\ }\bibfield  {title} {\enquote {\bibinfo {title} {Evaluating
  exponential integrators for {Markov} chain ion channel models},}\ }\href@noop
  {} {\bibfield  {journal} {\bibinfo  {journal} {Computing in Cardiology}\
  }\textbf {\bibinfo {volume} {42}},\ \bibinfo {pages} {885--888} (\bibinfo
  {year} {2015}{\natexlab{b}})}\BibitemShut {NoStop}%
\bibitem [{\citenamefont {Hinch}\ \emph {et~al.}(2004)\citenamefont {Hinch},
  \citenamefont {Greenstein}, \citenamefont {Tanskanen}, \citenamefont {Xu},\
  and\ \citenamefont {Winslow}}]{Hinch-etal-2004}%
  \BibitemOpen
  \bibfield  {author} {\bibinfo {author} {\bibfnamefont {R.}~\bibnamefont
  {Hinch}}, \bibinfo {author} {\bibfnamefont {J.~L.}\ \bibnamefont
  {Greenstein}}, \bibinfo {author} {\bibfnamefont {A.~J.}\ \bibnamefont
  {Tanskanen}}, \bibinfo {author} {\bibfnamefont {L.}~\bibnamefont {Xu}}, \
  and\ \bibinfo {author} {\bibfnamefont {R.~L.}\ \bibnamefont {Winslow}},\
  }\bibfield  {title} {\enquote {\bibinfo {title} {A simplified local control
  model of calcium-induced calcium release in cardiac ventricular myocytes},}\
  }\href@noop {} {\bibfield  {journal} {\bibinfo  {journal} {Biophys J}\
  }\textbf {\bibinfo {volume} {87}},\ \bibinfo {pages} {3723--3736} (\bibinfo
  {year} {2004})}\BibitemShut {NoStop}%
\bibitem [{\citenamefont {Tikhonov}(1952)}]{Tikhonov-1952}%
  \BibitemOpen
  \bibfield  {author} {\bibinfo {author} {\bibfnamefont {A.~N.}\ \bibnamefont
  {Tikhonov}},\ }\bibfield  {title} {\enquote {\bibinfo {title} {Systems of
  differential equations containing small parameters in the derivatives},}\
  }\href@noop {} {\bibfield  {journal} {\bibinfo  {journal} {Mat. Sb. (N.S.)}\
  }\textbf {\bibinfo {volume} {31(73)}},\ \bibinfo {pages} {575--586} (\bibinfo
  {year} {1952})}\BibitemShut {NoStop}%
\bibitem [{\citenamefont {Fenichel}(1979)}]{Fenichel-1979}%
  \BibitemOpen
  \bibfield  {author} {\bibinfo {author} {\bibfnamefont {N.}~\bibnamefont
  {Fenichel}},\ }\bibfield  {title} {\enquote {\bibinfo {title} {Geometric
  singular perturbation theory for ordinary differential equations},}\
  }\href@noop {} {\bibfield  {journal} {\bibinfo  {journal} {Journal of
  Differential Equations}\ }\textbf {\bibinfo {volume} {31}},\ \bibinfo {pages}
  {53--98} (\bibinfo {year} {1979})}\BibitemShut {NoStop}%
\bibitem [{\citenamefont {Biktashev}(2002)}]{Biktashev-2002-PRL}%
  \BibitemOpen
  \bibfield  {author} {\bibinfo {author} {\bibfnamefont {V.~N.}\ \bibnamefont
  {Biktashev}},\ }\bibfield  {title} {\enquote {\bibinfo {title} {Dissipation
  of the excitation wavefronts},}\ }\href@noop {} {\bibfield  {journal}
  {\bibinfo  {journal} {Phys Rev Lett}\ }\textbf {\bibinfo {volume} {89}},\
  \bibinfo {pages} {168102} (\bibinfo {year} {2002})}\BibitemShut {NoStop}%
\bibitem [{\citenamefont {Biktashev}\ and\ \citenamefont
  {Suckley}(2004{\natexlab{a}})}]{Biktashev-Suckley-2004-PRL}%
  \BibitemOpen
  \bibfield  {author} {\bibinfo {author} {\bibfnamefont {V.~N.}\ \bibnamefont
  {Biktashev}}\ and\ \bibinfo {author} {\bibfnamefont {R.}~\bibnamefont
  {Suckley}},\ }\bibfield  {title} {\enquote {\bibinfo {title} {Non-{Tikhonov}
  asymptotic properties of cardiac excitability},}\ }\href@noop {} {\bibfield
  {journal} {\bibinfo  {journal} {Phys Rev Lett}\ }\textbf {\bibinfo {volume}
  {93}},\ \bibinfo {pages} {168103} (\bibinfo {year}
  {2004}{\natexlab{a}})}\BibitemShut {NoStop}%
\bibitem [{\citenamefont {Biktasheva}\ \emph
  {et~al.}(2006{\natexlab{a}})\citenamefont {Biktasheva}, \citenamefont
  {Simitev}, \citenamefont {Suckley},\ and\ \citenamefont
  {Biktashev}}]{Biktasheva-etal-2006-PTRSA}%
  \BibitemOpen
  \bibfield  {author} {\bibinfo {author} {\bibfnamefont {I.~V.}\ \bibnamefont
  {Biktasheva}}, \bibinfo {author} {\bibfnamefont {R.~D.}\ \bibnamefont
  {Simitev}}, \bibinfo {author} {\bibfnamefont {R.~S.}\ \bibnamefont
  {Suckley}}, \ and\ \bibinfo {author} {\bibfnamefont {V.~N.}\ \bibnamefont
  {Biktashev}},\ }\bibfield  {title} {\enquote {\bibinfo {title} {Asymptotic
  properties of mathematical models of excitability},}\ }\href@noop {}
  {\bibfield  {journal} {\bibinfo  {journal} {Philos Transact Roy Soc A}\
  }\textbf {\bibinfo {volume} {364}},\ \bibinfo {pages} {1283--1298} (\bibinfo
  {year} {2006}{\natexlab{a}})}\BibitemShut {NoStop}%
\bibitem [{\citenamefont {Simitev}\ and\ \citenamefont
  {Biktashev}(2011)}]{Simitev-Biktashev-2011}%
  \BibitemOpen
  \bibfield  {author} {\bibinfo {author} {\bibfnamefont {R.~D.}\ \bibnamefont
  {Simitev}}\ and\ \bibinfo {author} {\bibfnamefont {V.~N.}\ \bibnamefont
  {Biktashev}},\ }\bibfield  {title} {\enquote {\bibinfo {title} {Asymptotics
  of conduction velocity restitution in models of electrical excitation in the
  heart},}\ }\href@noop {} {\bibfield  {journal} {\bibinfo  {journal} {Bull
  Math Biol}\ }\textbf {\bibinfo {volume} {73}},\ \bibinfo {pages} {72--115}
  (\bibinfo {year} {2011})}\BibitemShut {NoStop}%
\bibitem [{\citenamefont {Clancy}\ and\ \citenamefont
  {Rudy}(2002)}]{Clancy2002}%
  \BibitemOpen
  \bibfield  {author} {\bibinfo {author} {\bibfnamefont {C.~E.}\ \bibnamefont
  {Clancy}}\ and\ \bibinfo {author} {\bibfnamefont {Y.}~\bibnamefont {Rudy}},\
  }\bibfield  {title} {\enquote {\bibinfo {title} {{Na$^{+}$} channel mutation
  that causes both {B}rugada and long-{QT} syndrome phenotypes: a simulation
  study of mechanism.}}\ }\href@noop {} {\bibfield  {journal} {\bibinfo
  {journal} {Circulation}\ }\textbf {\bibinfo {volume} {105}},\ \bibinfo
  {pages} {1208--1213} (\bibinfo {year} {2002})}\BibitemShut {NoStop}%
\bibitem [{\citenamefont {Biktashev}(2003)}]{Biktashev2003}%
  \BibitemOpen
  \bibfield  {author} {\bibinfo {author} {\bibfnamefont {V.}~\bibnamefont
  {Biktashev}},\ }\bibfield  {title} {\enquote {\bibinfo {title} {Envelope
  equations for modulated non-conservative waves},}\ }\href@noop {} {\bibfield
  {journal} {\bibinfo  {journal} {IUTAM Symposium Asymptotics, Singularities
  and Homogenisation in Problems of Mechanics}\ }\textbf {\bibinfo {volume}
  {5(1)}},\ \bibinfo {pages} {11} (\bibinfo {year} {2003})}\BibitemShut
  {NoStop}%
\bibitem [{\citenamefont {Suckley}\ and\ \citenamefont
  {Biktashev}(2003)}]{Suckley2003}%
  \BibitemOpen
  \bibfield  {author} {\bibinfo {author} {\bibfnamefont {R.}~\bibnamefont
  {Suckley}}\ and\ \bibinfo {author} {\bibfnamefont {V.~N.}\ \bibnamefont
  {Biktashev}},\ }\bibfield  {title} {\enquote {\bibinfo {title} {Comparison of
  asymptotics of heart and nerve excitability.}}\ }\href@noop {} {\bibfield
  {journal} {\bibinfo  {journal} {Phys Rev E}\ }\textbf {\bibinfo {volume}
  {68}},\ \bibinfo {pages} {011902} (\bibinfo {year} {2003})}\BibitemShut
  {NoStop}%
\bibitem [{\citenamefont {Biktashev}\ and\ \citenamefont
  {Suckley}(2004{\natexlab{b}})}]{Biktashev2004}%
  \BibitemOpen
  \bibfield  {author} {\bibinfo {author} {\bibfnamefont {V.~N.}\ \bibnamefont
  {Biktashev}}\ and\ \bibinfo {author} {\bibfnamefont {R.}~\bibnamefont
  {Suckley}},\ }\bibfield  {title} {\enquote {\bibinfo {title} {Non-{T}ikhonov
  asymptotic properties of cardiac excitability.}}\ }\href@noop {} {\bibfield
  {journal} {\bibinfo  {journal} {Phys Rev Lett}\ }\textbf {\bibinfo {volume}
  {93}},\ \bibinfo {pages} {168103} (\bibinfo {year}
  {2004}{\natexlab{b}})}\BibitemShut {NoStop}%
\bibitem [{\citenamefont {Biktasheva}\ \emph
  {et~al.}(2006{\natexlab{b}})\citenamefont {Biktasheva}, \citenamefont
  {Simitev}, \citenamefont {Suckley},\ and\ \citenamefont
  {Biktashev}}]{Biktasheva2006}%
  \BibitemOpen
  \bibfield  {author} {\bibinfo {author} {\bibfnamefont {I.~V.}\ \bibnamefont
  {Biktasheva}}, \bibinfo {author} {\bibfnamefont {R.~D.}\ \bibnamefont
  {Simitev}}, \bibinfo {author} {\bibfnamefont {R.}~\bibnamefont {Suckley}}, \
  and\ \bibinfo {author} {\bibfnamefont {V.~N.}\ \bibnamefont {Biktashev}},\
  }\bibfield  {title} {\enquote {\bibinfo {title} {Asymptotic properties of
  mathematical models of excitability.}}\ }\href {\doibase
  10.1098/rsta.2006.1770} {\bibfield  {journal} {\bibinfo  {journal} {Philos
  Transact Roy Soc A}\ }\textbf {\bibinfo {volume} {364}},\ \bibinfo {pages}
  {1283--1298} (\bibinfo {year} {2006}{\natexlab{b}})}\BibitemShut {NoStop}%
\bibitem [{\citenamefont {Biktashev}\ \emph {et~al.}(2008)\citenamefont
  {Biktashev}, \citenamefont {Suckley}, \citenamefont {Elkin},\ and\
  \citenamefont {Simitev}}]{Biktashev2008}%
  \BibitemOpen
  \bibfield  {author} {\bibinfo {author} {\bibfnamefont {V.~N.}\ \bibnamefont
  {Biktashev}}, \bibinfo {author} {\bibfnamefont {R.}~\bibnamefont {Suckley}},
  \bibinfo {author} {\bibfnamefont {Y.~E.}\ \bibnamefont {Elkin}}, \ and\
  \bibinfo {author} {\bibfnamefont {R.~D.}\ \bibnamefont {Simitev}},\
  }\bibfield  {title} {\enquote {\bibinfo {title} {Asymptotic analysis and
  analytical solutions of a model of cardiac excitation.}}\ }\href {\doibase
  10.1007/s11538-007-9267-0} {\bibfield  {journal} {\bibinfo  {journal} {Bull
  Math Biol}\ }\textbf {\bibinfo {volume} {70}},\ \bibinfo {pages} {517--554}
  (\bibinfo {year} {2008})}\BibitemShut {NoStop}%
\bibitem [{Note1()}]{Note1}%
  \BibitemOpen
  \bibinfo {note} {Note that diagonalizability and reality of the eigenvalues
  of the full transition rate matrix ${\protect \mathaccentV {hat}05EA}$ can be
  guaranteed under the assumption of detailed balance~\cite {detailed-balance},
  and ${\protect \mathaccentV {hat}05EA_f}=\protect \qopname \relax
  m{lim}_{{\varepsilon }\to 0}\left ({\varepsilon }{\protect \mathaccentV
  {hat}05EA}\right )$.}\BibitemShut {Stop}%
\bibitem [{\citenamefont {West}\ \emph {et~al.}(2015)\citenamefont {West},
  \citenamefont {Bridge}, \citenamefont {White}, \citenamefont {Paszek},\ and\
  \citenamefont {Biktashev}}]{West-etal-2015-JMB}%
  \BibitemOpen
  \bibfield  {author} {\bibinfo {author} {\bibfnamefont {S.}~\bibnamefont
  {West}}, \bibinfo {author} {\bibfnamefont {L.~J.}\ \bibnamefont {Bridge}},
  \bibinfo {author} {\bibfnamefont {M.~R.~H.}\ \bibnamefont {White}}, \bibinfo
  {author} {\bibfnamefont {P.}~\bibnamefont {Paszek}}, \ and\ \bibinfo {author}
  {\bibfnamefont {V.~N.}\ \bibnamefont {Biktashev}},\ }\bibfield  {title}
  {\enquote {\bibinfo {title} {A method of `speed coefficients' for biochemical
  model reduction applied to the {NF-kappaB} system},}\ }\href {\doibase
  10.1007/s00285-014-0775-x} {\bibfield  {journal} {\bibinfo  {journal} {J Math
  Biol}\ }\textbf {\bibinfo {volume} {70}},\ \bibinfo {pages} {591--620}
  (\bibinfo {year} {2015})}\BibitemShut {NoStop}%
\bibitem [{\citenamefont {Star{\'y}}(2016)}]{Stary-2016}%
  \BibitemOpen
  \bibfield  {author} {\bibinfo {author} {\bibfnamefont {T.}~\bibnamefont
  {Star{\'y}}},\ }\emph {\bibinfo {title} {Mathematical and Computational Study
  of {Markovian} Models of Ion Channels in Cardiac Excitation}},\ \href@noop {}
  {Ph.D. thesis},\ \bibinfo  {school} {University of Exeter} (\bibinfo {year}
  {2016})\BibitemShut {NoStop}%
\bibitem [{\citenamefont {Anderson}(1991)}]{detailed-balance}%
  \BibitemOpen
  \bibfield  {author} {\bibinfo {author} {\bibfnamefont {W.~J.}\ \bibnamefont
  {Anderson}},\ }\href@noop {} {\emph {\bibinfo {title} {Continuous Time
  {Markov} Chains. An Application-Oriented Approach}}}\ (\bibinfo  {publisher}
  {Springer},\ \bibinfo {address} {New York etc},\ \bibinfo {year}
  {1991})\BibitemShut {NoStop}%
\end{thebibliography}
\providecommand{\noopsort}[1]{}\providecommand{\singleletter}[1]{#1}%

\end{document}